\documentclass[onecolumn,useAMS]{mn2e}
\usepackage{epsfig}
\usepackage{times} 
\def\la{\lower.5ex\hbox{$\; \buildrel < \over \sim \;$}}
\def\ga{\lower.5ex\hbox{$\; \buildrel > \over \sim \;$}}
\def\apj{ApJ}

\def\apjl{ApJL} 

\def\aj{AJ}

\def\physrep{Physics Reports}

\begin{document}

\title[HI Signal from Reionization Epoch]{HI Signal from Reionization Epoch}
\author[Shiv K. Sethi]
{Shiv K. Sethi$^1$  \\
\hspace{-0.1cm} ${}^1$Raman Research Institute, Bangalore 560080, India \\
\hspace{-0.1cm} emails: sethi@rri.res.in
}

\maketitle

\begin{abstract}
We investigate the all-sky signal in redshifted  atomic  hydrogen 
(HI) line from the 
reionization epoch.  We model the phase of reionization as multiple 
point sources which carve out spherical Stromgren spheres. We study 
ionization histories compatible with WMAP observation.  The Lyman-$\alpha$ 
and soft x-ray emission from these sources is taken into account for
studying the HI signal.  HI can be observed in both emission and absorption
depending on the ratio of Lyman-$\alpha$ to ionizing flux and the spectrum
of the radiation in soft xray. We also compute the signal from pre-reionization
epoch and show that within the uncertainty in cosmological parameters, it 
is fairly robust. The main features of  HI signal can be summarized as: (a)
The pre-ionized  HI can be seen in  absorption for 
$\nu \simeq 10\hbox{--}40 \, \rm MHz$; the maximum signal strength is 
$\simeq 70\hbox{--}100 \, \rm mK$.  (b) A sharp absorption
feature of width $\la 5 \, \rm MHz$ might be 
observed in the frequency  range $\simeq 50 \hbox{--}100 \, \rm MHz$,  
depending on the reionization history. The strength of the signal is 
 proportional  to  the ratio of   the Lyman-$\alpha$ and the 
hydrogen-ionizing flux and the spectral index of the radiation field 
in soft xray (c)  At larger frequencies, 
HI is seen in emission  with peak frequency  between
 $60\hbox{--}100 \, \rm MHz$, depending on the ionization history
of the universe; the  peak strength  of this signal is $\simeq  50 \, \rm mK$.
From Fisher matrix analysis, we compute  the precision with which the 
parameters of the model can be estimated from a future experiment: (a) 
the pre-reionization signal can constrain a region in the  
$\Omega_b h^2$--$\Omega_m h^2$ plane  (b) HI observed in emission can be 
used to give precise, $\la 1 \%$,  measurement of the  evolution of the 
ionization fraction in the universe, and (c) the transition region from
absorption to emission can be used as  a 
 probe of the spectrum of ionizing sources; in particular, the  HI signal in
this regime can give reasonably precise measurement of the fraction of the 
universe heated by soft x-ray photons. 

\end{abstract}

\section{Introduction}
One of the most important issues in modern cosmology is to 
understand the re-ionization of 
the universe. In the standard scenario the universe is mostly neutral
for $z \la 1000$ up to an epoch when the formation of first structures
 might have reionized the universe (e.g. Peebles 1993, Padmanabhan 2002, 1993).
 From Gunn-Peterson 
(GP) tests
in past decades the intergalactic medium (IGM) 
was inferred to be almost fully ionized for $z \la 5$ (see e.g. Barkana \& Loeb 2001,  Peebles 1993). Many recent 
observations  suggest that universe might have been making a transition from
fully ionized  to neutral for $5.5 \la z \la 6.2$ 
(White et~al. 2003, Fan et~al. 2002, Djorgovski et~al. 2001, Becker et~al. 2001). Recent WMAP observations
support a re-ionization epoch in the  redshift  range 
 $z \simeq 10\hbox{--}25$ depending on
the details of reionization  (Kogut et al. 2003). 
Together GP and 
WMAP observations probably point to a complex reionization history in which
the universe might have gone through two epochs of reionization (e.g. Mesinger \& Haiman 2004, Wyithe \& Loeb 2004b). Given
these uncertainties it is therefore important that the phase of reionization
is investigated using other probes. One such probe is based on 
observing the redshifted neutral hydrogen (HI) hyperfine  line 
 during reionization (see e.g. Scott \& Rees 1990; Shaver et al. 1997).  

The observable effect in this probe is the change in CMBR temperature at the 
redshifted frequency $1420/(1+z) \, \rm MHz$. Even in the pre-reionization
era this change is non-zero and potentially detectable (Scott \& Rees 1990). 
During reionization the signal is further complicated by Lyman-$\alpha$ and 
x-ray radiation fields (Madau, Meiksin, \& Rees 1997). 	Madau et al. (1997)
computed the expected HI signal around an ionizing source. More 
recently,  Gnedin and Shaver (2004) studied  the HI signal from hydrodynamic
simulations. In this 
paper we  attempt to model  the phase of reionization  semi-analytically to 
study the global HI signal. 
We model the phase of reionization as multiple ionizing sources, whose
individual spherical 
Stromgren spheres expand to percolate, causing  the universe to
 change from entirely neutral to a fully ionized phase 
(e.g. Haiman \& Holder 2003). We compute the 
background HI signal in this semi-analytic model, 
taking into account the effect of different radiation fields. We also 
compute the signal from pre-ionization epoch and show that  the 
present uncertainty in cosmological parameters are small enough 
to  fix quite robustly the main
features of this signal. 

In addition to the background signal it is possible to consider fluctuations
in the HI signal (e.g. Gnedin and Shaver 2004, Zaldarriaga, Furlanetto, \& Hernquist  2004, 
Tozzi et~al. 2000). 
Both single dish and interferometric experiments are being planned to 
detect the HI signal from the reionization epoch 
 (Subrahmanyan  2004, {\tt www.lofar.org},  Pen, Wu, \& Peterson 2004).  Future
interferometric experiments like LOFAR have the potential to detect 
both the all-sky and  the 
fluctuating component of the HI signal. In this paper we discuss only the 
all-sky  signal. This signal is harder to detect owing largely to 
galactic and extragalactic 
foregrounds which have temperatures several orders of magnitude 
larger than the expected HI signal.  However unlike the HI signal these 
foregrounds are expected to be featureless in frequency and  therefore 
might be removable (see e.g. Zaldarriaga et al. 2004 and references therein).  

In the next section we discuss the evolution of HI spin temperature in
an expanding universe. In \S 3 we discuss  the pre-ionization
signal. In \S 4 we discuss in detail the semi-analytic model for 
the ionization history and the effects  of Lyman-$\alpha$ and 
soft x-ray radiation fields in
determining the HI signal. 
 In \S 5 we present 
our main  results. In \S 6 we discuss various uncertainties 
associated with our model and summarize our results. 
 Throughout  this paper, unless specified,  we
use the currently-favoured FRW  model: spatially flat
with $\Omega_m = 0.3$ and $\Omega_\Lambda = 0.7$ (Spergel et al. 2003, Perlmutter  et al. 1999,
Riess  et al. 1998) with  $\Omega_b h^2 = 0.02$ (Spergel et al. 2003, Tytler
 et al. 2000) and
$h = 0.7$ (Freedman  et al. 2001).

\section{Evolution of HI spin temperature}
The HI  spin temperature $T_s$  is defined as:
\begin{equation}
{n_2 \over n_1} = 3 \exp(-T_\star/T_s)
\end{equation}
Here $n_2$ and $n_1$ are the populations of the hyperfine states.
$T_\star = h \nu_\star/k = 0.06 \, \rm K$; $\nu_\star = 1420 \, \rm MHz$ is the frequency of hyperfine transition. In the early universe, 
 the spin temperature 
 is determined from detailed balancing between various processes that
 can alter the 
relative populations of the two levels (Field 1958, 1959):
\begin{equation}
T_s = {T_{\rm CMBR} + y_c T_K + y_\alpha T_\alpha \over 1 +  y_c + y_\alpha}
\label{eqts}
\end{equation}
Here $y_c = C_{21}/A_{21} T_\star/T_K$,  $C_{21} \propto n_{\rm \scriptscriptstyle H}$ is the rate of collisional de-excitation of the 
upper level (Field 1958, Allison \& Dalgarno 1969); $y_\alpha = D_{21}/A_{21} T_\star/T_\alpha$,  
$D_{21} \propto n_{\alpha}$ is the  
de-excitation rate of the upper triplet induced by Lyman-$\alpha$ photons, with  $n_\alpha$ 
being the number density of 
Lyman-$\alpha$ photons (Field 1958). $y_c$ and $y_\alpha$  correspond, respectively, 
 to   relative 
probabilities with which 
collisions between atoms and the presence of Lyman-$\alpha$ photons  determine
the level populations.  $T_K$ is the matter temperature and $T_\alpha$ is the 
'temperature' of Lyman-$\alpha$ photons in the 
frequency range $\simeq \nu_\alpha \pm \nu_\star$ (Field 1959). Eq.~(\ref{eqts}) assumes that  CMBR is the only radio source in the universe. 
In the epoch
following the recombination and before the reionization starts this is
 evidently true; in our analysis we assume it to be the case
even after the start of 
re-ionization.  In the pre-reionization era, there are no Lyman-$\alpha$ photons, and therefore $y_\alpha = 0$.  During reionization Lyman-$\alpha$ photons 
can play an important role in determining HI hyperfine level population.
Eq.~(\ref{eqts}) is valid in the expanding universe so long as all the 
processes that determine the level population have rates far exceeding the 
expansion rate of the universe, and therefore 
the evolution of the spin temperature is determined by the slow 
expansion rate and is given by the values  of different
quantities at any given epoch; it can be shown that 
all the relevant rates are large enough for this
approximation to hold  (see e.g. Madau et al. 1997). 
 
As CMBR is the only radio source at high redshifts, 
HI signal in hyperfine transition can be seen against the 
 CMBR in emission or in  absorption. The observed quantity then is the 
deviation of CMBR from a black body at frequency $\simeq \nu_\star$. 
The observed deviation at the present epoch is, at the
observed frequency  $\nu_0 = \nu_\star/(1+z)$:
\begin{equation}
\Delta T_{\rm CMBR} = -{\tau_{\rm \scriptscriptstyle HI} \over (1+z)}(T_{\rm CMBR} -T_s)
\label{eqh1}
\end{equation}
Here $\tau_{\rm \scriptscriptstyle HI} = \sigma_\nu N_{\rm HI} T_\star/T_s$,
with the HI column density $N_{\rm HI} = \int n_{\rm \scriptscriptstyle HI} d\ell$; $\sigma_\nu = c^2 A_{21}\phi_\nu/(8\pi \nu_\star)$;  
 $A_{21} \simeq 1.8 \times 10^{-15} \, \rm sec^{-1}$ 
and $\phi_\nu$ is the line response function. For a flat response function,
$\phi_\nu = 1/\Delta\nu$, this can be expressed as (see e.g. Madau et al. 1997):
\begin{equation}
\tau_{\rm \scriptscriptstyle HI} = 0.02 \left ({\Omega_{\rm HI} h^2 \over 0.024} \right ) \left ({0.15 \over \Omega_m h^2} \right )^{1/2} \left ({ T_{\rm CMBR} \over T_s} \right ) \left ( {1+z \over 20} \right )^{1/2} 
\label{eqh1p}
\end{equation}
Here $\Omega_{\rm HI}$ corresponds to   the neutral fraction of hydrogen. Eqs.~(\ref{eqh1})~and~(\ref{eqh1p}) can readily be extended to the case when the universe
 is a mixture of phases with different ionized fraction and spin temperature.
We shall see later that it turns out to be the case when the universe makes
a transition from a fully neutral to a fully ionized phase. 
It follows from  Eq.~(\ref{eqh1}) that  the  HI can be observed in either 
absorption or emission against CMBR depending on whether $T_s$ is less 
than or exceeds $T_{\rm CMBR}$.

\section{HI signal from pre-reionization epoch}
The pre-reionization signal has been computed previously by 
Scott and  Rees (1990) and more recently by Loeb and Zaldarriga (2004). 
We re-compute and summarize this signal in this section. In particular
we emphasize the robustness of this signal and its dependence of 
this signal on various cosmological parameters.

To calculate this signal we compute the evolution of matter temperature 
and ionized fraction of hydrogen in the post-recombination prior to
the epoch of reionization (for relevant equations see e.g. Peebles 1993).
To accurately determine the dependence of ionization fraction in the
 post-recombination universe on cosmological parameters,
 we begin integrating the equations at $z = 2000$. At this 
redshift the hydrogen is fully ionized, and the helium, $8\%$ of the baryon
number density,
can be considered fully neutral. 

 At $z \simeq 1000$, $T_{\rm CMBR} = T_K$ and 
it follows from Eq.~(\ref{eqts}) that $\Delta T_{\rm CMBR} = 0$ (Scott 
\& Rees 1990). 
For $z \la  100$, $T_K < T_{\rm CMBR}$ and therefore HI can 
be observed in absorption against CMBR if $y_c \ga 1$ (Eq.~(\ref{eqh1})). 
In Figure~\ref{fig:f2} we  show the  pre-ionization signal as a function of the observing 
frequency, $\nu_0 = \nu_\star/(1+z)$ for a universe that doesn't 
re-ionize for $z \ga 10$. As discussed above, for $z \gg 100$, 
$T_s \simeq T_{\rm CMBR}$ and therefore the HI can be observed neither
in emission nor in absorption. For $z \la 100$, $y_c > 1$ and therefore
$T_s \simeq T_K < T_{\rm CMBR}$; the HI can be observed in absorption against
the CMBR. For $z \la 20$, $y_c < 1$ and therefore $T_s \simeq T_{\rm CMBR}$
and $\Delta T_{\rm CMBR} \simeq 0$. Scott and Rees (1990) 
showed these main  features of the absorption signal. However owing to 
uncertainties in the cosmological parameters like $\Omega_B$ and $\Omega_m$ 
neither the magnitude nor the frequency at which it will peak could 
be established. In recent year fairly precise determination of these
cosmological parameters has become possible largely owing to
CMBR anisotropy and large scale structure studies (e.g Spergel et~al. 2003). 

 The main dependence of the absorption
signal is on different cosmological parameters that determine the evolution 
of the matter temperature $T_K$ and the probability of collisions $y_c$. 
In Figure~\ref{fig:f2} we also show the dependence of the HI signal on
 different 
cosmological parameters;
 the parameter range is taken to be $2\sigma$ around the 
central values and errors determined by WMAP (Spergel et~al. 2003). It is 
seen from the figure that the location of the signal in frequency space and 
also its magnitude are fairly robust within the expected uncertainties
in determining cosmological parameters. As we shall see below re-ionization
at $z \simeq 15$ doesn't affect the main features of the absorption signal. 
This signal therefore is a unique probe of the thermal history of
 the universe.  In a later section, we show how the detection of this 
signal can be used to determine cosmological parameters.

\section{HI signal during re-ionization}
In the standard $\rm \Lambda CDM$ model of structure formation in the 
universe, small inhomogeneities, which originated during an
inflationary era in the very early universe,  were amplified by
gravitational instability. First structures in the universe formed 
when these perturbations became non-linear and collapsed (see e.g. Peebles 1993, Padmanabhan 2002, 1993 and references therein). 
The gravitational collapse 
of these structures
might  set off star-formation  (alternatively some material might 
end up in black holes which by accreting more matter will radiate  with
harder spectrum than first star-forming galaxies, see e.g. Ricotti \& Ostriker 2003) which will emit 
UV light and ionize the IGM. The process of re-ionization of the 
universe is generally quite complicated and not well understood (e.g. Barkana \& Loeb 2001). However 
it is possible to study it  within the framework of simple 
models which might  give important clues about the details of this 
process. We assume that each collapsed object emits isotropically and 
causes a spherical ionization sphere around it. The reionization is 
completed when the fraction of volume occupied by these 
ionized spheres approaches unity.  Important ingredients of this problem are:
 (a) Halo population
at high redshift, (b) molecular and atomic cooling in Haloes, (c) Initial 
Mass Function of stars and star formation rate, (d) Escape fraction of 
UV  photons from Haloes, (e) clumpiness of the IGM. Of these the most uncertain
are (c) and (d) and have to be modelled using simple parameterized models. 
In addition to these uncertainties in modelling the ionization fraction of 
the universe, other  considerations are needed to understand the HI signal
during reionization. These complications can be appreciated by
studying the signal around an isolated object. The ionizing radiation
from the object will carve out a spherical ionizing front. The region 
outside this ionized sphere
 is mostly neutral and its HI signal can  be observed. The object however
emits not only the hydrogen-ionizing radiation but can emit radiation
close to Lyman-$\alpha$ frequency and also in soft x-ray. These two radiation
fields can penetrate beyond the ionized sphere and alter the level
population of the HI. If the Lyman-$\alpha$ intensity is high enough 
the level population of HI is coupled to Lyman-$\alpha$ photons (Eq.~(\ref{eqts})). In addition Lyman-$\alpha$ photons can slowly heat the HI gas by
atom recoil (Madau et~al. 1997). The soft x-ray photons can heat the 
gas beyond the ionized sphere by photo-electric absorption (Madau et~al. 1997).
As we shall study below, the region immediately beyond the ionized sphere is 
heated by soft x-ray  photons 
to temperatures above $T_{\rm CMBR}$ and therefore will be seen in
emission (Eq.~\ref{eqh1}). There are two possibilities 
for regions not heated by soft x-ray: 
(a)   the spin temperature in these regions is   $\simeq T_{\rm CMBR}$, 
and therefore  these regions cannot be observed in either emission or absorption. (b)  the spin temperature $T_s$ in these regions  is determined 
by the  Lyman-$\alpha$ photons which means that 
$T_s \simeq T_\alpha \simeq T_K$.  As $T_K \la T_{\rm CMBR}$ in regions not yet heated by soft x-ray photons, the 
 HI from these regions can be observed in absorption. As we shall see later,
possibility (b) is realized in most situations. 

\subsection{Evolution of Ionized fraction}
We take up various aspects of the evolution of the ionized component in this
subsection. 

 The first important ingredient is the mass function of 
the collapsed dark matter haloes. In the $\rm \Lambda CDM$ models, the 
mass function  at any redshift can be obtained from the Press-Schechter
 method (Press \& Schechter 1974, for details see 
 Pebbles 1993, Padmanabhan 2002, 1993). The number
density of dark matter haloes per unit mass is: 
\begin{equation}
{dn \over dM} = \sqrt{{2\over\pi}}{\rho_m \over M}\delta_c(z) \left | {d\sigma \over dM} \right | {1 \over \sigma^2(M)} \exp\left [-\delta_c(z)^2/(2\sigma^2(M)) \right ]
\label{psme}
\end{equation}
Here $\sigma(M)$ is the mass dispersion filtered at the scale corresponding 
to mass $M$ in the linear theory; $\sigma(M)$ for length scale 
corresponding to $8 h^{-1} \, \rm  Mpc$ is $\simeq 0.9$ (Spergel et~al. 2003); $\delta_c(z) \simeq 1.7 D(0)/D(z)$, 
with $D(z)$ being the solution of growing mode in linear theory; in sCDM model
$D(0)/D(z) = (1+z)$ (for details see Padmanabhan 2002, 1993, Peebles 1980). 
The  collapsed fraction of all 
structures can be calculated from Eq.~(\ref{psme}); for $z \simeq 20$
the collapsed fraction is $\simeq 2 \times 10^{-3}$.
Even though dark matter haloes of all masses will collapse and 
virialize,  they will not be able to trap baryons unless
the haloes have masses exceeding the Jeans mass of the 
IGM. The Jeans mass at $z  \simeq 20$ is $\simeq 10^4 \, \rm M_\odot$ 
(e.g. Barkana \& Loeb 2001).  
Another important consideration is whether the baryons 
in the collapsed haloes  can cool sufficiently 
rapidly to form stars (see e.g. Tegmark et~al. 1997, 
Sethi 2004, Barkana \& Loeb 2001);
 the smallest halo mass that satisfy this
criterion is $M \simeq 10^7 \, \rm M_\odot$. Our study 
shows that the evolution of ionization
fraction  (Eq.~(\ref{eq:ionfra}) is insensitive to the exact value of the mass of the 
 smallest  halo  that 
can form stars.  Another complication that is harder to incorporate in our
analysis is that the Jeans mass in the regions which are ionized is 
very different from the regions which are neutral. In the ionized regions,
$T_K\simeq 10^4 \, \rm K$ and  at $z \simeq 20$, this gives a Jeans mass 
$\simeq 10^9 \rm M_\odot$. This is not an important consideration 
 so long as the ionized fraction $ f_{\rm ion} \ll 1$ (Eq.~(\ref{eq:ionfra})).
 However once the  ionized fraction approaches unity, the mass of the smallest
haloe that can cool sufficiently to 
form stars should be comparable to the value of 
Jeans mass in the ionized region. We check how this  affects 
the evolution of ionized fraction
  by retaining only haloes larger than the Jeans
mass at that redshift 
 in the ionized regions as the ionized fraction exceeds 0.5 and 
find that it makes an insignificant difference to the evolution of 
ionized fraction. 

From information about the halo population and cooling arguments it is 
possible to speculate on the ionization history of the universe from
photo-ionization. The main uncertainty is the hydrogen-ionizing luminosity
(and its evolution) of each halo, which has to  be parameterized. Assuming 
that  halo of mass $M$ emits isotropically the hydrogen-ionizing luminosity
$\dot N_\gamma$ (in photons~$\rm sec^{-1}$), the radius of ionizing sphere 
around the source will satisfy the equation (Stromgren Sphere, see e.g. 
Shu 1992, Shapiro \& Giroux 1987, Wyithe \& Loeb 2004b):  
\begin{equation}
{dR \over dt} -H R={(\dot N_\gamma - 4\pi/3 R^3 \alpha_B C  n_b^2 x_{\rm \scriptscriptstyle HI})  \over (\dot N_\gamma + 4 \pi R^2 x_{\rm \scriptscriptstyle HI} n_b)}
\label{stromsp}
\end{equation}
Here $C\equiv \langle n_b^2 \rangle/\langle n_b \rangle^2$ is the clumping 
factor of the IGM. In Eq.~(\ref{stromsp}) the right hand side approaches 
unity  as $R$ tends to zero. This corresponds  to the fact that initially
Stromgren sphere expands at the speed of light (Wyithe \& Loeb 2004b).
 At larger $R$, the $\dot N_\gamma$ term in the denominator can be dropped
 and the Stromgren sphere 
evolution equation 
 approaches it usual form (e.g. Shapiro \& Giroux 1987).     
Using Eq.~(\ref{psme}), the fraction of the universe that
 is ionized at a given redshift is (see e.g. Haiman \& Holder 2003):
\begin{equation}
f_{\rm ion}(z) = {4 \pi \over 3}\int_{0}^z dz' \int dM {dn \over dM}(M,z') R^3(M,z,z')
\label{eq:ionfra}
\end{equation}
 The  minimum 
mass that can contribute to Eq.~(\ref{eq:ionfra}) is taken to be 
$10^7 \, \rm M_\odot$. As discussed above the
 evolution  ionized fraction is not very sensitive 
to the mass of smallest halo that can form stars.
Further assuming that the luminosity of the source is $\propto M$, the 
ionized fraction can be calculated in terms of the evolution of the photon
luminosity of a single halo of some fiducial mass and the evolution of 
the clumping factor. For simplicity we assume the clumping factor to have 
a constant value between one and five, and take the luminosity  of 
the  halo to be constant. Therefore, the  photon luminosity of a halo of a given mass at 
any redshift can be expressed as:
\begin{equation}
\dot N_\gamma(M,t) = \left ({M \over 5 \times 10^7 \, \rm M_\odot } \right )\dot N_\gamma(0).
\end{equation}
 In Figure~\ref{fig:f3} we show several
ionization histories for different values of $\dot N_\gamma(0)$
 and clumping factor $C$.  
All the ionization
histories shown in Figure~\ref{fig:f3} are consistent with WMAP observations which
suggest that $\tau_{\rm reion} = 0.17 \pm 4$, 68\% confidence level
 (Kogut et~al. 2003) : the 
dot-dashed curve corresponds to 
$\tau_{\rm reion} =  0.22$ if the universe remains fully ionized
 after the first reionization; the 
dashed line gives, $\tau_{\rm reion} = 0.13$ . 
 It appears that $ \dot N_\gamma \simeq 10^{50}$ is required 
to ionize the universe early enough to satisfy WMAP observations. This is 
just three orders of magnitude below the photon luminosity of a typical
star-burst galaxy (see .e.g. Leitherer et~al. 1999). 
Alternatively one can infer  that the efficiency of the first star
formation was very high (see e.g. Haiman \& Holder 2003, Chiu, Fan \& 
Ostriker 2003). There are other 
uncertainties like feedback from supernova, and photo-dissociation of 
molecular hydrogen which indicate this estimate is  a lower limit 
(see e.g. Loeb \& Barkana 2001, 
 Haiman \& Holder  2003, Haiman,  Rees, \& Loeb 1997,  Dekel \& Silk 1986). 
Alternatively it is possible that the collapsed fraction of the universe
far exceeded the value given by $\rm \Lambda CDM$ models and was caused 
by some other physical process like tangled magnetic fields
 (Sethi \& Subramanian 2005).

\subsection{Evolution of HI signal}
\subsubsection{Effect of Lyman-$\alpha$  radiation}
As discussed above the HI signal during re-ionization 
depends on the Lyman-$\alpha$ and x-ray flux from the objects. While the 
photons with frequencies around the ionization threshold are absorbed inside
the Stromgren sphere, photons with frequencies between Lyman-$\alpha$ 
($1216 \, \rm  \AA$)  and 
the ionization threshold ($912 \, \rm  \AA$) in the rest frame of the object
free-stream  into the medium beyond the Stromgren sphere. (More exactly
only photons between Lyman-$\alpha$ and Lyman-$\beta$ frequencies 
can free-stream; higher frequency photons are absorbed locally by 
resonant transitions to higher levels (Lyman-$\beta$ or above). This 
however makes no essential change to our inference here.) Assuming 
 Lyman-$\alpha$ response function to be Dirac delta function, a photon
of frequency $\nu$ between Lyman-$\alpha$ and the ionization threshold is 
absorbed by the resonant scattering at a comoving distance 
$\Delta R \simeq H_0^{-1} \Omega_m^{-1/2}(1+z)^{-3/2} \Delta \nu/\nu_\alpha$
 with  $\Delta \nu = \nu - \nu_\alpha$ in the expanding universe; 
$\Delta R \simeq 30 h^{-1} \, \rm Mpc$ for $\nu \simeq 13.6 \rm eV$. This 
defines the region of influence of the 'Lyman-$\alpha$' photons around a 
source. Similar to ionized fraction, the fraction of the universe 
influenced by Lyman-$\alpha$ photons can be defined as:
\begin{equation}
f_{\alpha}(z) = {4 \pi \over 3}\int_{0}^z dz' \int dM {dn \over dM}(M,z') R_\alpha^3(z,z')
\end{equation}
with $R_\alpha \simeq H_0^{-1}\Omega_m^{-1/2}(1+z)^{-5/2} \Delta z$ with 
$\Delta z = z' -z$.   It can be readily seen 
 that $f_\alpha$ rapidly reaches unity.
 In light of this fact, it 
is not necessary to study the region of influence around each source; 
instead it is possible to study the evolution of the mean intensity of the 
Lyman-$\alpha$ photons. The mean proper specific intensity of 
Lyman-$\alpha$ photons  at high redshifts is given by:
\begin{equation}
I_{\nu_\alpha}(z) \simeq  {H_0^{-1} hc(1+z)^3 \over 4\pi \Omega_m^{1/2}}\int dM\int_z^{z_{\rm max}}dz' {dn \over dM}\dot N_{\gamma \alpha}(M,z') (1+z')^{-5/2-\beta} 
\end{equation}
Here $\dot N_{\gamma \alpha}$ is the photon luminosity in Lyman-$\alpha$ photons of any object and $\beta$ is the spectral index of the photons in the 
frequency range between  Lyman-$\alpha$ and the Lyman continuum, $\nu_0$. Here
 $1+ z_{\rm max} = (1+z)\nu_0/\nu_\alpha$ and is determined from the fact that photons above 
Hydrogen ionization threshold are absorbed locally. Given the uncertainty 
about the spectrum of the ionizing sources,  we assume
$\beta = 0$ and $\dot N_{\gamma \alpha} = A \dot N_\gamma$ i.e.
 the 
'Lyman-$\alpha$' luminosity (.i.e. the photon luminosity of the source
in the wavelength range between Lyman-$\alpha$ and Lyman-$\beta$ in the 
rest frame of the object) is a constant multiple of the luminosity of 
the  Hydrogen-ionizing 
photons.  From
Eq.~(\ref{eqts}) it follows that the HI  level population can be determined 
by the Lyman-$\alpha$ photons if $n_\alpha = 4\pi I_{\nu_\alpha}/(hc)$ 
is high enough to give $y_\alpha \ga  T_{\rm  CMBR}/T_\alpha$. In the medium 
in which HI couples to Lyman-$\alpha$, $T_\alpha$ tends to relax to 
$T_K$  near the line center (Field 1959). 
Detailed calculations for an expanding universe show that this can 
occur in a time-scale shorter than the expansion time scale of 
the  universe (Rybicki \& Dell'antonio 1994).  We will
 use $T_\alpha = T_K$  throughout this paper. In addition 
Lyman-$\alpha$ photons can also lead to 
heating of the HI from atom recoil (Field 1959, Madau et~al. 1997). This
heating rate however is generally quite small (Chen \& Miralda-Escud\'e 2003) 
and we neglect it in our analysis. In Figure~\ref{fig:f4} and~\ref{fig:f5} we show the influence 
of Lyman-$\alpha$ radiation on the spin temperature. The figures show that
for most models we consider the Lyman-$\alpha$ flux is large enough to 
couple spin temperature to the matter temperature for $z \ga 20$. This result
is in qualitative agreement with the analysis of Ciardi \& Madau (2003).  

\subsubsection{X-ray Heating Outside Stromgren Sphere}
 To determine the 
the thermal evolution of HI during reionization we also need to take into
account the heating of the HI from soft x-ray photons. If the ionizing 
sources emit soft x-ray photons, then these photons can reach beyond the 
Stromgren sphere and heat the surrounding HI from photo-electric absorption.  
The heating rate from the soft x-ray photons at a distance $R$ from an
ionizing source is:
\begin{equation}
 \Delta \dot E_{\rm x} = f {\dot N_\gamma h \over 4 \pi R^2} \int_{\nu_0}^{\infty} d\nu \nu^{-1-\alpha}
\nu_0^\alpha
\left [(\nu - \nu_0) \sigma_\nu^{\rm \scriptscriptstyle HI} + \chi (\nu - \nu_1) \sigma_\nu^{\rm \scriptscriptstyle HeI} \right ] \exp(-\tau)
\label{heatxr}
\end{equation}
Here \{$\nu_0, \nu_1 = 13.6, 24.6 \, \rm eV$ \} are the ionization threshold
of Hydrogen and Neutral Helium, respectively; $f$, the fraction of 
soft x-ray photon energy deposited  for heating the medium is a sensitive 
function of the ionized  fraction, rising from nearly $0.1$ for an ionized 
fraction $\simeq 2 \times 10^{-4}$, the residual ionized fraction 
from recombination,  to nearly one for the fully ionized medium  
 (Shull \& van Steenberg 1985). Soft x-ray photons will also partially 
ionize the medium (see below), thereby raising the ionized fraction above the 
value from recombination. The fraction of 
energy deposited in reionizing HI fall rapidly above the ionization fraction
$\ga 0.1$ (Shull \& van Steenberg 1985). Therefore, without going into
the details of the reionization from soft xray 
 it is safe to assume that  the ionization fraction remains  $\la 0.1$, 
as verified by the study of Venkatesan et al. (2001). For the class of 
model we study, we find that 
the ionization fraction  
due to  reionization from  soft x-ray photons varies from roughly $0.1$ just
outside the Stromgren sphere to 
  $10^{-3}$  up to  radii at  which
the x-ray heating can cause appreciable heating (for details see below). The 
fraction of heat deposited in this range of ionization fraction is 
$0.1 \la f  \la 0.3$. 
As this is hard to self-consistently take 
into account at each radius,  we use a constant, average, 
 value $f = 0.2$  throughout this paper.

  The heating due to ionization of singly 
ionized helium is negligible and is neglected here. $\chi = 0.08$ is the 
primordial Helium to Hydrogen ratio. We have assumed the spectrum of 
photons above hydrogen ionization  threshold to fall as: $\nu^{-\alpha}$.  $\sigma_\nu^{\rm \scriptscriptstyle HI}$ and $\sigma_\nu^{\rm \scriptscriptstyle HeI}$ 
are the cross-sections for Hydrogen and Helium ionization, respectively. 
$\tau = (n_{\rm \scriptscriptstyle HI} \sigma_\nu^{\rm \scriptscriptstyle HI}  +  n_{\rm \scriptscriptstyle HeI} \sigma_\nu^{\rm \scriptscriptstyle HeI})R$ 
is the optical depth to hydrogen and Helium ionization. In Eq.~(\ref{heatxr})
we use the Newtonian expression for solving flux from a given source. This 
is justified as the size of the region influenced by the source $\ll H^{-1}(z)$.  Eq.~(\ref{heatxr}) 
is exact, however to solve it  simplifications are needed. 
Firstly the ionization level of different species is required to 
 self-consistently solve for the rate of energy injection. This however
 can be simplified 
by taking the gas to be fully ionized inside the Stromgren sphere
and neutral outside the Stromgren sphere. This is justified as 
even though the soft x-ray photons 
will  partially ionize the region outside the Stromgren sphere
   the ionization fraction
is generally so small that the region can be considered totally neutral
and  therefore only the heating effect of these photons needs to be considered
(for details see Venkatesan, Giroux  \& Shull  2001).
Therefore we can take
$R = r_s + r$, where $r_s$ is the radius of Stromgren sphere and can be 
solved for each source separately. This simplification allows us to
find the rate of energy injection from soft x-ray photons 
 at a distance $R$ from
any source.  As seen from Eq.~(\ref{heatxr}),  much of the contribution to
this rate come from radii $R \la 1/(n_{\rm \scriptscriptstyle HI} \sigma_\nu^{\rm \scriptscriptstyle HI})$; as $\sigma_\nu^{\rm \scriptscriptstyle HI}) \propto \nu^{-3}$ above the Hydrogen threshold, the region of influence of photons
for  different frequencies can be found. Eq.~(\ref{heatxr}) is needed
 to address
whether the matter temperature can be raised sufficiently above the 
CMBR temperature in this region of influence. 

The evolution of matter temperature outside the
 Stromgren sphere can be found by including the energy injection from 
 soft x-ray photons. However, from 
Eq.~(\ref{eqh1}) we notice that once $T_s \gg T_{\rm CMBR}$ the HI emission
is independent of the $T_s$ or from Eq.~(\ref{eqts}) the matter temperature. 
Therefore for our study we do not need to solve for the detailed matter
temperature  profile outside a source, we only need to know the radius of 
the region in which $T_s \gg T_{\rm CMBR}$. This is ensured if 
$T_k \gg T_{\rm CMBR}$ and $y_\alpha \ga 1$. The latter requirement 
is met in all the cases we consider.  Therefore  we calculate this radius by 
the following requirement: It is the radius in which
 the energy injection can raise the matter 
temperature  $T_K = q T_{\rm CMBR}$, with $q \ga  1$,  
in the local Hubble time, i.e. it is calculated from the expression:
\begin{equation}
{\Delta \dot E_{\rm x}(R,z) \over H(z) k_{\scriptscriptstyle B}} = q T_{\rm CMBR}
\label{xrad}
\end{equation}
This allows us to define the fraction of volume which is influenced by x-ray,
$f_{\rm x}$, 
similar to the ionization fraction. In Figure~\ref{fig:f6}, we show the evolution of 
$f_{\rm x}$ for several values of the spectral index $\alpha$ and $q$. 
$q \simeq 1$ roughly determines the radius which demarcates the region seen in 
absorption from the region seen in emission. It should be pointed out that
this  prescription always gives an overestimate of the signal. This 
is because the HI signal in the transition region from absorption to 
emission is smaller than  assuming  a sharp boundary between the two 
regions. However this uncertainty is much smaller
than the uncertainty in Lyman-$\alpha$ flux and the spectrum of the 
radiation field in  soft xray, as seen in Figure~\ref{fig:f6}. While
presenting our results in the next section, unless specified otherwise,
  we use $q = 1$ throughout. In Figure~\ref{fig:f6}, 
 the dependence of $f_{\rm x}$ on the spectral index $\alpha$  is seen 
to be quite strong ;  the observed signal, therefore,  
 will hold the promise of 
unravelling the nature of ionizing sources. 

In previous studies the the effect of x-ray heating in HI signal was 
taken into account by evolving 
 the average specific intensity of the x-ray background (Chen \& Miralda-Escude 2004). It would be possible to do it only if $f_{\rm x} \gg 1$. 
However, our study shows that, in contrast to Lyman-$\alpha$ photons, $f_{\rm x} \la 1$ for a large range of redshifts of interest
 (Figure~\ref{fig:f6}), i.e. 
the 'graininess' 
of the ionizing sources is important for x-ray heating. Even though both
these approaches give similar results for the all-sky HI signal, the
underlying physics is quantitatively quite different. In our study
the HI signal doesn't make transition from 
absorption to emission by an increase in the average x-ray specific intensity 
but by coalescence of regions in which x-ray heating can raise the 
temperature of matter above CMBR temperature.
In particular, the 
fluctuating component of the signal will be very different in the two cases.
In the first case, the only scale in the problem is the size of the  Stromgren 
sphere and the fluctuating component can be calculated by using methods
such as developed by Zaldarriaga et al. (2004). In the latter case, the region
affected by the x-ray around each source introduces a new scale in the 
problem and the fluctuating component is likely to be dominated by the 
gradients across the boundary of the transition region  between absorption
and emission. As future missions like LOFAR will have capability to 
observe in the redshift range where such effects might dominate,
 it is important to 
compute the fluctuating component of the HI signal for this multi-scale 
problem.

In this and the previous subsection, we have modelled the effect of 
Lyman-$\alpha$ and soft x-ray radiation in terms of two parameters: $A$ and
$\alpha$. It could be asked what the different values of these parameter
imply about the nature of sources of reionization. Given
the uncertainty of star-formation history/quasar formation rate, the IMF,
and metallicity of the sources at large redshifts, it is difficult to
predict the spectrum of ionizing sources at those redshifts. This 
uncertainty motivates us to consider simple models parameterizable in 
terms of two parameters. 
 
One possible way to speculate about the properties of these sources
at high redshift is by analogy with low redshift sources. 
Observed quasars 
have  a harder intrinsic spectrum $\alpha \simeq 1$  than is expected
of star-forming 
galaxies for which $\alpha \la 2.5$ 
(see e.g. Miralda-Escude \& Ostriker 1990). Therefore by varying the value of 
$\alpha$ one could model sources of either kind. The value of $A$ is very
hard to estimate from  observations at high redshifts.
 This is owing to the 
fact that much of flux short-ward of the Lyman-$\alpha$ is  likely 
to be removed from scattering in the intervening IGM
(see e.g. Barkana \& Loeb  2001 and references therein). It should be
pointed out that for this reason 
 direct observation of ionizing sources  by future telescopes 
like NGST also cannot reveal their nature. And  the only possible
way to estimate  e.g. Lyman-$\alpha$ flux of these sources might be  by its
effect on the HI signal.

To summarize the discussion of the previous two subsections: 
main features of the HI signal during the reionization around 
a single ionizing source  are: (a) In a  
region outside the
Stromgren sphere, x-ray  heating  raises the temperature
of the matter above CMBR temperature (Madau et~al. 1997).
 If the Lyman-$\alpha$ radiation in this region is large
enough, i.e. $y_\alpha \ga 1$, 
 to couple HI spin temperature to Lyman-$\alpha$ photons, then
using $T_\alpha = T_K$, the HI in this region can be seen in emission. If the 
Lyman-$\alpha$ flux is not large enough, then HI continues to be coupled to 
CMBR and therefore cannot be observed in emission or absorption against CMBR.
In all cases of interest $y_\alpha \ga 1$ and therefore
HI is seen in emission from this region.  
(b) photons between Lyman-$\alpha$ and Lyman-$\beta$ frequencies 
 penetrate deeper than the soft  x-ray photons and therefore there are 
regions that are affected by Lyman-$\alpha$ photons 
but are not heated by x-ray.  This  leads
to the possibility that while HI gets coupled the Lyman-$\alpha$ photons,
 the matter
temperature is still lower than CMBR temperature. In this case this 
region will be seen in absorption against the CMBR. 
We have  extended  this analysis to multiple sources
distributed randomly throughout the universe which is valid 
so long as  $f_{\rm ion} \la  1$ 
or the individual Stromgren spheres do not intersect. (A stronger condition
for the validity of such an analysis might be $f_{\rm x} \la  1$;  however it 
only means that once  $f_{\rm x} \simeq 1$ 
all the HI in the universe can only be observed in emission.) 

\section{Results}
As discussed above and shown in Figure~\ref{fig:f2}, the pre-ionization HI signal
is fairly robust and only depends on the thermal history of the universe. 
The HI signal during re-ionization however is 
fairly complicated; Figures~\ref{fig:f3}~to~\ref{fig:f6} capture this
 uncertainty in 
modelling the ionization  and thermal history of the regions 
outside the ionizing region. Generically, 
the HI is seen in emission during reionization  from
the regions heated to temperature in excess of 
 CMBR temperature from soft xray or it 
can be observed in absorption from  region which are colder than
CMBR  but HI is coupled 
to Lyman-$\alpha$ radiation. The all-sky  HI signal 
is a linear combination
of these two components:
\begin{equation}
\Delta T_{\rm CMBR} = -{\tau_{\rm \scriptscriptstyle HI}^{\rm em} \over (1+z)}(T_{\rm CMBR} -T_s^{\rm em})-{\tau_{\rm \scriptscriptstyle HI}^{\rm absor} \over (1+z)}(T_{\rm CMBR} -T_s^{\rm absor})
\label{eqh1_a}
\end{equation}
As discussed in the previous section, the spin temperature in the 
region seen in absorption $T_s^{\rm absor} \simeq T_{\rm K}$ and 
the spin temperature in the region see in emission $T_s^{\rm em} \gg T_{\rm CMBR}$; $\tau_{\rm \scriptscriptstyle HI}^{\rm em} 
\propto  (f_{\rm x} - f_{\rm ion})$ and 
$\tau_{\rm \scriptscriptstyle HI}^{\rm absor}  \propto
 (1 - f_{\rm x})$; 
$\max\{f_{\rm x}, f_{\rm ion}\} = 1$. Using Eq.~(\ref{eqh1p}),  Eq.~(\ref{eqh1_a}) simplifies to:
\begin{equation}
\Delta T_{\rm CMBR} = 0.06 \, {\rm K}  \left [(f_{\rm x} - f_{\rm ion}) - (1-f_{\rm x})\left (1 -{T_s^{\rm absor} \over T_{\rm CMBR}} \right ) \right ] 
\left ({\Omega_{\rm b} h^2 \over 0.024} \right ) \left ({0.15 \over \Omega_m h^2} \right )^{1/2} \left ( {1+z \over 20} \right )^{1/2}
\label{finsig} 
\end{equation}

In Figure~\ref{fig:f7}, we show the HI signal for several ionization histories
compatible with WMAP results; the 
dependence of the signal on the Lyman-$\alpha$ flux and the spectral
index of radiation in soft xray is also shown. The main
features of the HI signal are: (a) HI can be observed in absorption for 
$\nu \simeq 10\hbox{--}40 \, \rm MHz$; the maximum signal strength is 
$\simeq 70\hbox{--}100 \, \rm mK$.  This is the  signal from the 
pre-reionization epoch and as shown in Figure~\ref{fig:f2} is quite 
insensitive 
to the details of reionization. (b) A sharp absorption
feature of width $\la 5 \, \rm MHz$ might be 
observed in the frequency  range $\simeq 50 \hbox{--}100 \, \rm MHz$ 
depending on the reionization history. The strength of the signal is 
correlated 
with the ratio of   the Lyman-$\alpha$ and the 
hydrogen-ionizing flux and the soft xray spectral index.
   (c)  HI is seen in emission 
after the entire medium
is heated to temperatures above $T_{\rm CMBR}$. This signal peaks between
frequencies $60\hbox{--}100 \, \rm MHz$ depending on the ionization history
of the universe and has peak strength $\la 50 \, \rm mK$. Our results
are in qualitative agreement with the results of Gnedin \& Shaver (2004). 
However the average signal we obtain is roughly a factor of two larger
than obtained by Gnedin \& Shaver (2004). We can partly understand 
it by comparing 
the ionization histories in our models with the fiducial ionization models
 studied by them (Figure~4 in Gnedin \& Shaver (2004)). In the fiducial
models they study, the ionization fraction has  values $\ga 0.5$
for $z \simeq 20$. In our ionization models,  shown in Figure~\ref{fig:f3}, 
 the ionization
fraction increases at a slower rate, and reaches values roughly $\la 0.1$
by the epoch all the HI  in the universe is heated  above 
$T_{\rm CMBR}$ by the x-ray radiation. 

\subsection{Parameter estimation from HI signal}
The HI signal shown in Figure~\ref{fig:f7} can be used to estimate parameters
 of 
the underlying models. We first list all the parameters used in our 
analysis: Cosmological parameters, $\Omega_m$, $\Omega_b$ and $h$; parameters
related to modelling ionization history: photon luminosity
$\dot N_\gamma(0)$, clumping factor $C$; additional parameter needed
for modelling HI signal: the ratio of Lyman-$\alpha$ to hydrogen-ionizing 
luminosity, $A$, and the spectral index of ionizing radiation in soft x-ray 
wave-band. This is the minimum set of parameters required to 
study the HI signal. 
It should be pointed out that using these parameters we attempt
to model a physical process which might be inherently non-parameterizable.
For instance we assume all the evolution of ionizing 
luminosity to be the same for all sources. However
it is conceivable that all sources have different evolution 
and it is not possible in this 
analysis or even a more detailed simulation, to take that into account in
a parametric form. However,  we show in this section that 
it is  possible to glean some generic 
properties of the HI signal in the parametric analysis. 

HI signal can roughly be broken into three separate regimes: pre-reionization
signal, transition from absorption to emission, HI observed in emission. 
We discuss each of these regimes in detail below and argue how these 
different regimes can be used to extract a sub-set of the parameters. 

\subsubsection{ Regime 1: The pre-reionization signal}
The all-sky 
HI signal in this regime doesn't 
depend on the details of  reionization, i.e. in Eq.~(\ref{finsig}) $f_{\rm ion} = f_{\rm x} = 0$,  but  only on the 
cosmological parameters of the background universe. Before embarking on
a detailed statistical analysis of this signal, we discuss the 
dependence of this signal on the cosmological parameters. Eq.~(\ref{finsig})
shows the dependence of this signal on $\Omega_b h^2$ and $\Omega_m h^2$. 
In addition, from Eq.~(\ref{eqts}), the spin temperature, $T_s$ is seen
to depend on $n_b \propto \Omega_b h^2$ and on the evolution of matter 
temperature. The evolution of matter temperature, $T_K$ is determined 
by: (a) the local expansion rate. At large redshifts the 
expansion rate  $H^2 \propto \Omega_m h^2$, independent of the dark energy 
content or the geometry of the universe
 and (b) the fraction of 
ionized hydrogen in the post-recombination universe, $x_e$; 
in the post-recombination universe prior to reionization $x_e \propto 
(\Omega_m^{0.5} h/\Omega_b h^2)$ (e.g. Peebles 1993). To summarize this 
discussion, the HI signal in the pre-reionization epoch depends on two
parameters $\Omega_m h^2$ and $\Omega_b h^2$. Both these parameters 
can therefore be extracted from the observed HI signal. Future missions like
LOFAR can potentially detect the HI signal for $\nu \ga 20 \, \rm  MHz$ with 
RMS sensitivity $S_{\rm rms} \simeq 1 \, \rm mK$ (the primary aim 
of LOFAR is to measure the fluctuating component of the signal; however 
the fluctuating component can be detected against a smoothly varying mean 
signal (Zaldarriaga et al. 2004), and therefore the mean is also potentially
measurable).  A similar sensitivity is being 
aimed for the single dish experiments which plan to detect only the 
all-sky signal (Subrahmanyan  2004). From Figure~\ref{fig:f7},
 the pre-reionization signal
is seen to be the dominant signal for $\nu \la 40 \, \rm MHz$. 
We model the observed 
signal as six frequency  bins separated by $4 \, \rm MHz$  
 between $20 \, \rm MHz$ and $40 \, \rm MHz$ with signal determined to 
an RMS accuracy  $S_{\rm rms}$ in each bin (the signal as seen from Eq.~(\ref{finsig})  doesn't depend on the bin width). We perform Fisher matrix 
analysis for  the HI signal given by the central values of parameters
$\Omega_m h^2 = 0.14$ and $\Omega_bh^2 = 0.024$, as determined by WMAP, 
to extract information about these parameters. The Fisher matrix 
can be written  as (for a detailed discussion on Fisher matrix 
see e.g. Tegmark, Taylor  \& Heavens 1997):
\begin{equation}
{\cal F}_{ij}  = \sum_k{\partial (\Delta T_{\rm CMBR}(\nu_k)) \over \partial \theta_i}{1 \over S_{\rm rms}^2} {\partial (\Delta T_{\rm CMBR}(\nu_k)) \over \partial \theta_j}
\label{fishmat}
\end{equation}
Here $\theta_i = \{\Omega_b h^2, \Omega_m h^2 \}$ are the parameters to 
be determined and $k$ signifies summing over the frequency bins.

 In Figure~7 we show the 2-$\sigma$ contours for the parameters for two
 assumed values of  $S_{\rm rms}$. As seen from the figure, 
none of the two parameters
are well determined but a line of degeneracy is picked (we confirm this 
from principal component analysis of the Fisher matrix (see .e.g. Bond
\& Efstathiou 1999)),   which suggests that
what is well determined is some weighted ratio of the two parameters. This 
is owing to the fact that the parameter dependence is dominated by 
the pre-factor of Eq.~(\ref{finsig}) which gives such a ratio.  

How does the determination of parameters $\Omega_b h^2$ and $\Omega_m h^2$ 
compare with other measurements? $\Omega_b h^2$ can be determined to
 a reasonably precision
 from nucleosynthesis arguments (e.g. Tytler et al. 2000) and is 
the best determined parameter from the CMBR anisotropy measurements (Spergel et al. 2003,  Efstathiou \& Bond 1999); recent WMAP results give: $\Omega_b h^2 = 0.024 \pm 0.001$ (1$\sigma$).  $\Omega_m h^2  \simeq  0.14 \pm 0.02$ (1$\sigma$) from
recent WMAP observation and is determined largely  by the early 
integrated Sachs-Wolfe  effect (e.g  Efstathiou \& Bond 1999).
Also,  the  galaxy power spectrum 
depends on $\Omega_m h^2$ with a weak dependence on $\Omega_b$ (e.g. Bond 1996, Peebles 1993). Therefore, the  analysis of the observed galaxy power spectrum 
at $z \la 0.5$  gives independent information on this 
parameter which can be used  in conjunction with WMAP 
data to give tighter bounds
on parameters (e.g Tegmark et al. 2004). Figure~7 shows that it not possible to give
meaningful bounds on both $\Omega_m h^2$ and $\Omega_b h^2$. However,
 if we use 
prior information on one of the parameters  then the other parameter
can be determined with reasonable precision. For instance if we use the 
prior, given by $1\sigma$ error on the parameter, provided by WMAP data 
 on the value of $\Omega_b h^2$, then we 
obtain $\Omega_m h^2 \simeq 0.15 \pm 0.03$ (1$\sigma$) for $S_{\rm rms} = 1 \, \rm mk$, which is comparable to the 
precision of independent measurement of this parameter from WMAP. Therefore 
the pre-reionization HI signal could give important complementary information
on cosmological parameters. It should be pointed out that the
information on $\Omega_m h^2$, 
unlike the information from CMBR anisotropies  and galaxy surveys, is 
completely independent of the geometry or the dark energy content 
 of the universe, as it comes from 
the local expansion rate which is not affected by the geometry of the universe
or the dark energy  at $z \simeq 35$. 

\subsubsection{Regime 2: HI seen in emission} 
As seen in Figure~6, this part of the 
signal, assumed to last from the maximum of the signal to the 
frequency at which the universe is fully ionized, 
 can vary substantially depending on the ionization history of the 
universe. For the two ionization histories shown in Figure~\ref{fig:f7}, 
 the signal is observable from $\simeq 60 \, \rm MHz$ to $\simeq 100 \, \rm MHz$ or from  $\simeq 80 \, \rm MHz$ to $\simeq 120 \, \rm MHz$. In this 
regime also, the signal assumes reasonably simple form as $f_{\rm x} \ga 1$, or
all the  HI in the universe is observable  in emission; also in this 
regime, the signal   doesn't 
depend on  the  Lyman-$\alpha$ flux as $T_s \simeq T_K \gg T_{\rm CMBR}$, 
and therefore  the HI signal is nearly independent of $T_s$.
The main dependence of the HI signal in this regime is on the cosmological
parameters earlier discussed and the ionized fraction of  the universe. 
The ionized fraction is determined by photon luminosity $\dot N_\gamma(0)$,
the clumping factor $C$ and  $\Omega_m h^2$ and $\Omega_b h^2$, through
the matter power spectrum used to determine the abundance of haloes  
(e.g Bond 1996).

 First we consider the evolution of a single Stromgren sphere.  During 
this evolution, the second term 
on the right hand side of  Eq.~(\ref{stromsp}) is generally sub-dominant
 (Shapiro \& Giroux 1987), and therefore the dependence of
 the ionized fraction on
the clumping factor and the baryon density is generally weak, though
not negligible if the clumping factor is large, 
as see from Figure~3. In most cases, the evolution of a single Stromgren sphere is 
largely determined by the ratio: $\dot N_\gamma(0)/(\Omega_m h^2)$. 

 The other dependence of the HI signal 
on $\Omega_m h^2$ comes from  the matter power spectrum
 that determines the abundance of dark matter
haloes (e.g. Bond 1996, Eq.~(\ref{psme})). In addition, 
the pre-factor of Eq.~(\ref{finsig}) give the dependence on $\Omega_b h^2$ and 
$\Omega_m h^2$.

For Fisher matrix analysis we consider the following three 
parameters: the photon
luminosity, $\dot N_\gamma(0)$, $\Omega_m h^2$, and $\Omega_b h^2$. 
For this analysis  we take the HI signal to correspond to: 
$\dot N_\gamma(0) = 10^{49} \, \rm sec^{-1}$, $C = 1$ and $\Omega_bh^2 = 0.024$and $\Omega_m h^2 = 0.14$, the central values of determined by WMAP.  
Further,  we assume 
$S{\rm rms} = 1 \, \rm mK$ and 10 bins separated by $4 \, \rm MHz$ in the
frequency range $84 \, \rm MHz$ to $120 \, \rm MHz$. 

The principal component analysis of the Fisher matrix yields the 
following ratios of the eigenvalues: $\lambda_1:\lambda_2:\lambda_3 = \{1,0.02,
3 \times 10^{-8}\}$, with the eigenvectors, in the order of decreasing 
eigenvalues, dominated by $\Omega_b h^2$, $\Omega_m h^2$ and
 $\dot N_\gamma(0)$. The $1\sigma$ errors on parameters are:
$\Delta \Omega_b h^2/\Omega_b h^2 = 0.04$,  $\Delta \Omega_m h^2/\Omega_m h^2 = 0.07$ and 
$\Delta \dot N_\gamma(0)/\dot N_\gamma(0)  = 0.8$. 
 Therefore even though the cosmological parameters 
are well determined, the photon luminosity is poorly determined. This can be 
understood in terms of the evolution of a single Stromgren sphere. The 
dynamics of a single Stromgen sphere (Eq.~(\ref{stromsp})) is determined 
by the ratio $\dot N_\gamma(0)/(\Omega_m h^2)$ which causes this 
degeneracy. It should be noted that this degeneracy is generic to any
model which envisages the reionization of the universe in terms of 
expanding Stromgen spheres. However, it is possible to  break the
degeneracy by using prior information on 
 the cosmological parameters, as, as discussed above,  the 
cosmological parameters can be determined to high precision by 
other measurements like CMBR anisotropy. Using the prior on the cosmological 
parameters as $1\sigma$ error bars from WMAP measurement, the error on 
$\dot N_\gamma(0)$ is nearly halved. 
 In Figure~8 we show 
the contour levels in  $\dot N_\gamma(0)\hbox{--}(\Omega_m h^2)$ plane 
for $\Omega_bh^2$ fixed to the central value of WMAP. From Figure~8 it is 
seen that the photon luminosity, $\dot N_\gamma(0)$ 
 can be determined to an accuracy of around 
$50\%$.  Future CMBR experiment Planck  will measure the 
cosmological parameters
to a much higher precision (e.g. Efstathiou \& Bond 1999). Using the 
expected $1\sigma$ error bars for Planck measurements  as priors, 
assumed here for simplicity to be 
a factor of 5 smaller than the WMAP errors, 
the photon luminosity, $\dot N_\gamma(0)$  can be determined to  better than
$20\%$ relative precision. 

In the foregoing we attempt to determine  the evolution of ionization fraction
in terms of the physical parameters like photon luminosity. In this case the 
ionized fraction in all bins is correlated as it is determined by the same 
underlying physical parameters. Instead one could try to directly determine
the ionization fraction in each observable bin, independent of the underlying
physical model.

 In Fisher matrix analysis
 the ionization fraction in each bin can be treated as an 
independent  parameter to be 
determined .i.e. $\theta_i = \{f_{\rm ion}(\nu_i) \}$, here $\nu_i$ gives 
the central frequencies of the bins. In our analysis we therefore 
consider 12 parameters: $\{\theta_1, \theta_2 \} = \{\Omega bh^2, \Omega_mh^2\} $ and $\theta_i(i = 3,\cdots 12) = \{f_{\rm ion}(\nu_i) \}$. Owing to
the linearly of the HI signal on the ionized fraction, the derivative of 
the signal with respect to parameters corresponding to the ionized fraction in
each bin 
is rendered diagonal: 
\begin{equation}
{\partial (\Delta T_{\rm CMBR}(\nu_k)) \over \partial \theta_i} \propto \delta_{ik}
\end{equation}
 Here  $\delta_{ij}$ is the Kronecker delta function, and $i$ is in the  range 
between  $3$ and  $12$.
The Fisher matrix  reduces to:
 \begin{equation}
{\cal F}_{ij}  = \sum_k{\partial (\Delta T_{\rm CMBR}(\nu_k)) \over \partial \theta_i}{1 \over S_{\rm rms}^2} {\partial (\Delta T_{\rm CMBR}(\nu_k)) \over \partial \theta_j} P_{ij}
\label{fishmat1}
\end{equation} 
Here $P_{ij} = \delta_{ij}$  for 
 $i$ and $j$ both between from $3$ to $12$, and $P_{ij} = 1$ otherwise. Using the 
priors on cosmological parameters from WMAP, $S_{\rm rms} = 1 \, \rm mk$ 
and the model for HI signal as in the previous case, 
the relative precision on each parameter is: 
$\Delta \theta_i/\theta_i \la  0.01$ .i.e. the ionized fraction
 can be estimated be a precision of nearly $1 \%$.  It should be pointed 
out that for the model of HI signal we use 
 the ionization fraction increases from 
$0.15$ to $1$ in the frequency range we consider.  These errors are comparable 
to the errors one  would obtain by fixing the values of cosmological parameters
to the central value given by WMAP; in which case the Fisher matrix is 
diagonal. And the errors on the ionization fraction is simply the square-root
of the inverse of the diagonal elements. We also did the same analysis for 
different choices of parameters  and arrive at similar results.

To summarize this discussion: we discuss two possible way of extracting the 
ionization history of the universe from the observed HI signal. In one case 
one could attempt to estimate the underlying physical parameters like the 
photon luminosity of the halo of  some fiducial mass. On the other hand, it 
is conceivable that the ionization history is very complicated
 (e.g. Gnedin \& Shaver (2004) discuss many  complicated 
reionization histories) and it may not  be possible to readily extract the 
underlying physical parameters.
 In that case one could  try to directly estimate the ionization fraction 
in each observable bin. We show that it is possible to estimate the ionization 
fraction to a relative precision $\simeq 1 \%$ (for ionization fraction 
exceeding roughly $0.15$ in each bin)  from future experiments.
 This is comparable to 
the  determination of the  ionization history from future CMBR experiments
 (Kaplinghat et al. 2003). 

\subsubsection{Regime 3: Transition from absorption to emission}
 The HI signal in this 
regime carries information about the spectrum of the ionizing sources. 
We have modelled the spectrum of sources by two parameters: the ratio
of Lyman-$\alpha$ to hydrogen-ionizing luminosity, $A$ and the spectrum of 
ionizing radiation in soft-xray radiation, $\alpha$. The main observable 
features in this regime, as seen in Figure~6, 
 are a possible narrow trough in the frequency range $50\, \rm MHz$ to $80 \, \rm MHz$ and the rise to the maximum of the HI signal in emission. As seen in
Figure~6, the amplitude of the HI trough  is quite sensitive 
to the two parameters we use to model the spectrum of ionizing sources. The 
ionization fraction in this regime is $f_{\rm ion} \la 0.1$ and therefore 
not very important to understand the main features of the signal. 

 As in the previous
two cases, we can perform a Fisher matrix analysis for an assumed HI 
signal. The choice of parameters  to get maximum information, as
 discussed in the previous case, is important. It is possible to 
extract information about $A$ and $\alpha$ or alternatively, as in the 
previous discussion, one could directly attempt to measure $f_{\rm x}$ in
each bin. We adopt the latter approach as it would also be applicable to 
a more complicated signal. 

For our analysis, we take the underlying ionization model to be the 
same as in the previous case with $A = 20$ and $\alpha = 1.5$. 
 To take into account the narrowness of the 
features in this regime we consider 25 bin, each $1 \, \rm MHz$ wide in the 
frequency range $55 \, \rm MHz$ to $80 \, \rm MHz$. We consider 26 parameters:
$A$ and the values of $f_{\rm x}$ in each observable bin: $\theta_i(i = 2,\cdots 26) = \{f_{\rm x}(\nu_i) \}$; the cosmological parameters are assumed 
to be fixed to the central values determined by WMAP.  As discussed above,
the Fisher matrix in this case takes a simple form as the signal is linear
in $f_{\rm x}$ which give $25$ of the $26$ assumed parameters (Eq.~\ref{fishmat1}).   For $S_{\rm rms} = 1 \, \rm mK$, all  the 
principal components of the Fisher matrix  are well estimated (except the 
one corresponding to the frequency  at which the HI signal vanishes). 
From these components we show that: $\Delta A/A \simeq 0.1$. 
The percentage errors on the $f_{\rm x}$ are shown in Figure~9. The 
error on $A$ is quite sensitive to the assumed value of $A$ and 
$\alpha$. For instance 
for $A \la 2$ and $\alpha \la 3$, 
no meaningful bound can be obtained on $A$. In other words, 
the value of $A$  cannot be determined unless a trough  in HI signal is
 observed. The determination of $f_{\rm x}$ in each bin for 
$f_{\rm x} \ga 0.25$,  however, is 
less sensitive to the underlying model. We test this with models 
with different cosmological parameters and $A$ and find that the fractional
errors shown in Figure~9 are fairly indicative in these cases also.

In this subsection, we have shown how the observed HI signal can be used 
to get information about cosmological parameters and the ionizing sources. 
Our analysis allows us to glean the important information about the 
underlying parameters   in each of the three
regimes we consider.  We  attempt to estimate parameters we use in the 
modelling of the HI signal, like cosmological parameters, 
$\dot N_\gamma(0)$ and $A$. We   also
discuss methods which could give generic 
information  like the ionization
fraction or  the fraction of universe heated  by soft x-ray photons in 
an observable frequency bin. These methods would also be 
applicable  to  a more complicated HI signal.

\section{Discussion}
In the foregoing sections, we discussed the ramifications of various 
simplifying assumptions we make. In this section we
discuss many other effects of the our assumptions. In our analysis 
we do not take into effect the clustering of ionizing sources. The 
first objects to collapse in $\rm \Lambda CDM$ model 
 are  typically 3.5$\sigma$ peaks of the density
field (e.g. Barkana \& Loeb 2001), and consequently their centers 
would have been 
 more  clustered  than the density field 
by a factor of $\simeq 10$ (Kaiser 1984, for details see 
 Peebles 1993, Padmanabhan 1993 and references therein).
 This is important 
for studying the fluctuating component of the HI signal (e.g. Wyithe \& Loeb 20004a).  However, this  makes insignificant 
difference to  $f_{\rm ion}$ 
because the volume of the Stromgren sphere
scales roughly with the luminosity of the source (e.g. Madau et al. 1997).
 The effect of clustering 
on $f_{\rm x}$ is harder to access. While the energy injection (Eq.~({\ref{heatxr})) is proportional to the luminosity of the source, the radius, determined 
from Eq.~(\ref{xrad}), doesn't lend itself to some obvious scaling with 
the source luminosity. However, it is possible to draw  some 
general conclusions. The radius is largely determined by the exponent 
in  Eq.~({\ref{heatxr}) which is independent of the source luminosity. The 
effect of clustering would be a smaller number of 'ionizing centers' 
 with higher luminosities, and therefore $f_{\rm x}$ is expected to be smaller
in this case. This uncertainty in $f_{\rm x}$, though important, is 
likely to be within the uncertainty from the spectrum of radiation in the 
soft xray (Figure~6). Therefore the suite of models we present in Figure~7 
give a realistic picture of the evolution of the HI signal. 

It is possible to consider more complicated models of the ionizing 
sources (e.g. Haiman \& Holder 2003, Sethi 2004). It is conceivable that
each ionizing source underwent one episode of star-burst activity 
and most of the ionizing photons  were emitted during this epoch. Such
models do not give qualitatively new ionizing histories (e.g. Sethi 2004);
and the main features of the  HI signal are unlikely to be affected by it. Here
we only attempt to model the first phase of reionization, as required 
by the WMAP observation. Recent GP observations suggest the universe 
might have gone through two phases  of reionization (Mesinger \& Haiman 2004,
Wyithe \& Loeb 2004b). 
While this might change the HI signal for $z \simeq 6$, it is unlikely
to affect the main features of the HI signal from higher redshifts we 
consider here. 

Shaver et~al. (1998) considered the detectability of the all-sky HI signal
and concluded that signal-to-noise is not an important issue  in this
measurement. At observed frequencies 
 $\simeq 50 \, \rm MHz$, the system temperature is expected to 
be $\ga 1000 \, \rm K$, for a bandwidth of $\simeq 1 \, \rm MHz$, a 5-$\sigma$ 
detection is possible in an integration time of several hours. The main 
difficulties  are   system calibration and galactic and 
extra-galactic foregrounds. Galactic and extra-galactic foreground are 
expected to be smooth in frequency space and therefore potentially removable 
(Shaver et~al. 1998, Zaldarriaga et~al. 2004, Gnedin \& Shaver 2004
 and references therein). 
Calibration issues are being addressed  in on-going searches for the all-sky
signal (Subrahmanyan  2004).

\section*{Acknowledgment}
We  would like to thank Jayaram Chengalur, K. S. Dwarakanath,  Zoltan 
Haiman, Biman Nath,
 Anish Roshi,  Kandaswamy Subramanian, and Ravi Subrahmanyan for 
 for many useful discussions. We also thank Zoltan Haiman,  Ravi Subrahmanyan,
and Biman Nath for detailed comments on the manuscript.

\newpage

\begin{figure}
\epsfig{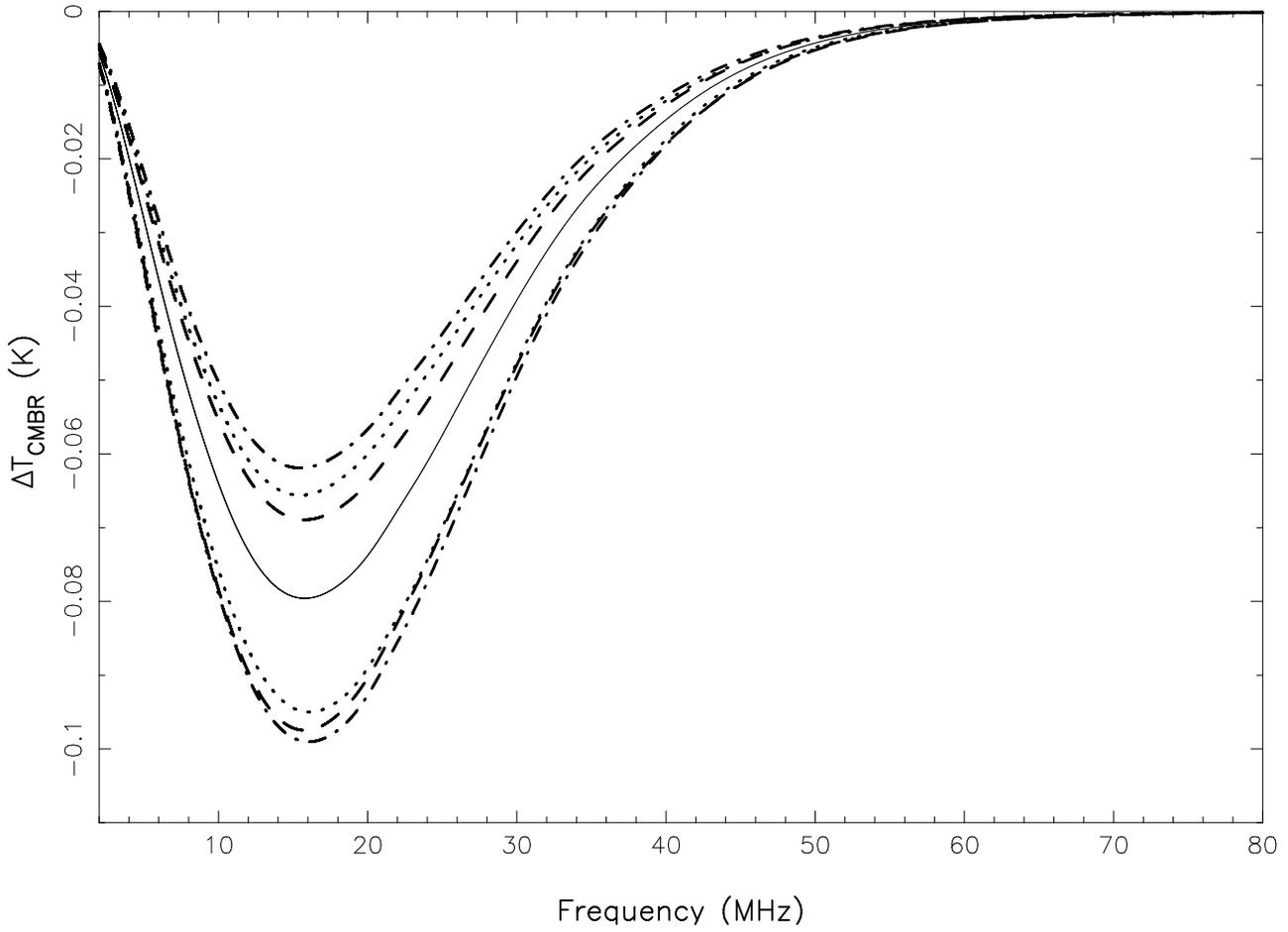}
\caption{The pre-ionization HI signal is shown as a function of 
observing frequency. The solid line corresponds to the best fit parameters
from WMAP: $\Omega_m = 0.3$, $\Omega_b = 0.047$, $h = 0.72$. Dotted, dashed, 
and dot-dashed lines give the 2-$\sigma$ envelope from WMAP 
observations for $\Omega_m$, $\Omega_b$ and $h$, respectively, 
around the best-fit model}
\label{fig:f2}
\end{figure}

\begin{figure}
\epsfig{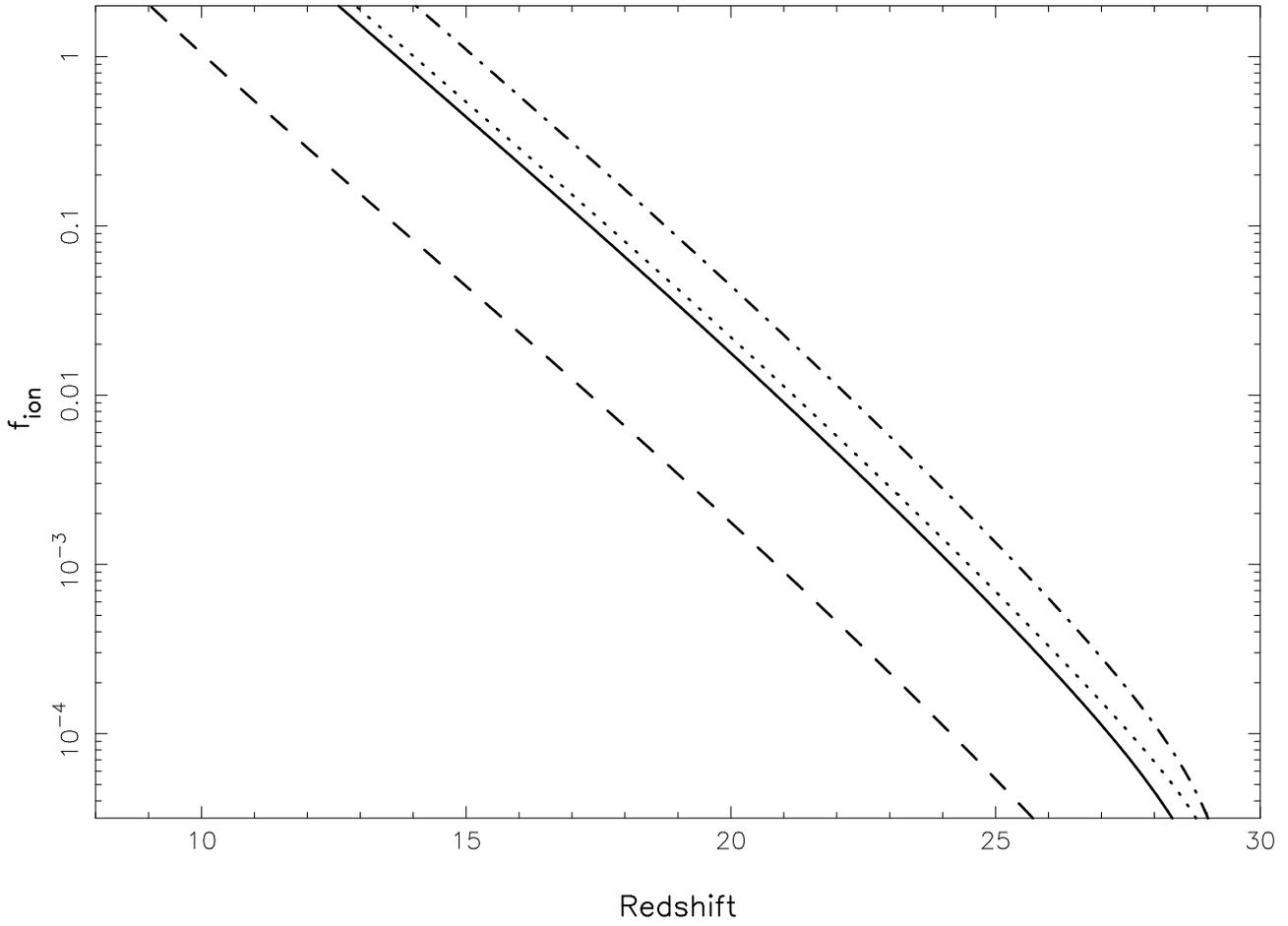}
\caption{The evolution of ionized fraction is shown for several 
models: $\dot N_\gamma(0) = 10^{50} \, \rm sec^{-1}$, $C = 1$ (Solid line), $\dot N_\gamma(0) = 3 \times 10^{50} \, \rm sec^{-1}$, $C = 1$ (dot-dashed line), $\dot N_\gamma(0) = 10^{49} \, \rm sec^{-1}$, $C = 1$ (dashed line),$\dot N_\gamma(0) = 3 \times 10^{50} \, \rm sec^{-1}$, $C = 4$ (dotted line)}
\label{fig:f3}
\end{figure}

\begin{figure}
\epsfig{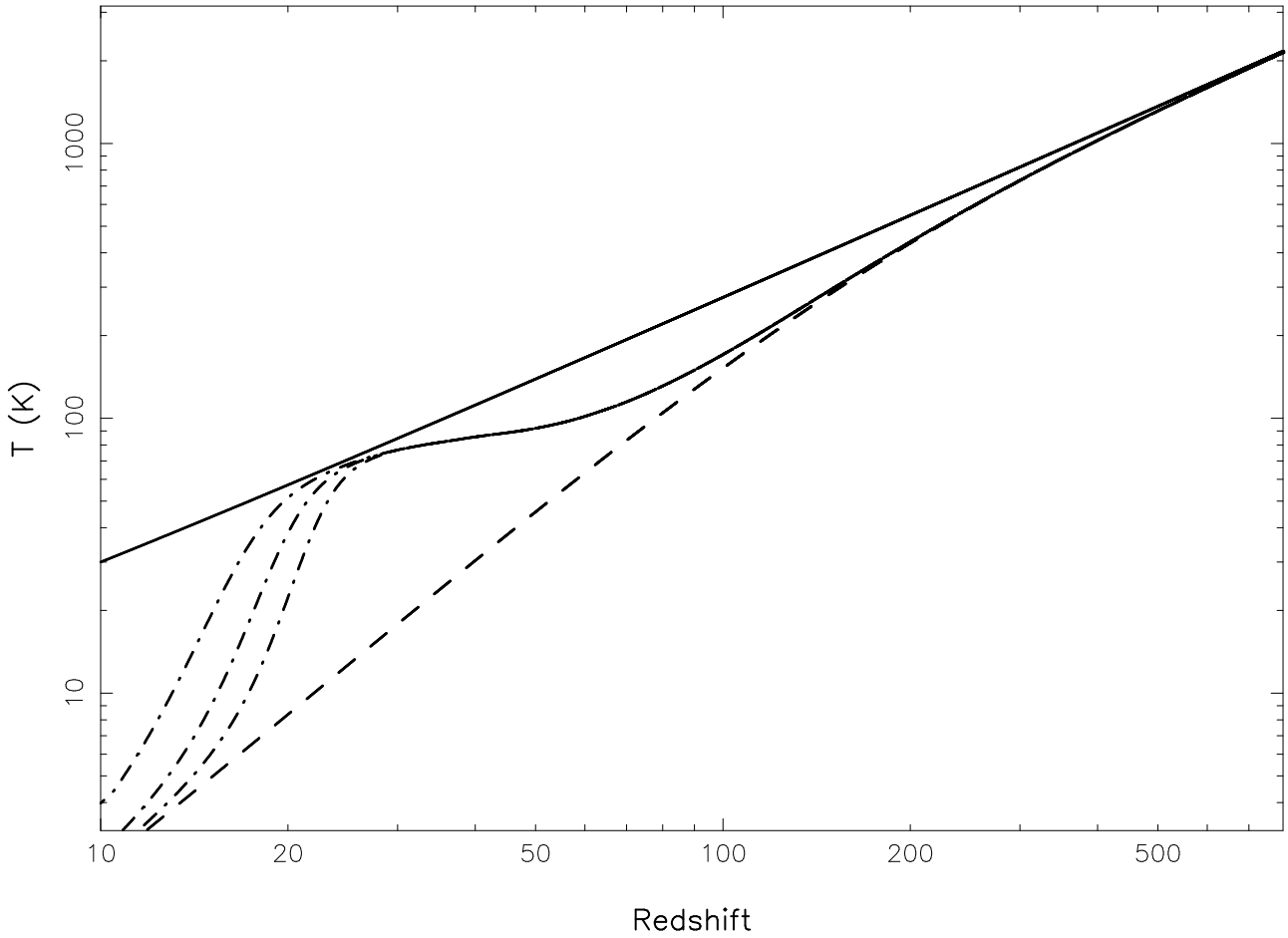}
\caption{ The evolution of spin temperature is shown for regions of the 
universe affected by Lyman-$\alpha$ radiation but not yet heated by the 
 x-ray. 
The three dot-dashed lines correspond, from bottom to top, to a 
re-ionization model
with $\dot N_\gamma(0) = 10^{50} \, \rm sec^{-1}$, and the ratio of Lyman-$\alpha$ to 
ionizing flux $A = \{40, 10, 2\}$, 
respectively. The solid and dashed lines correspond to CMBR and 
matter temperature, respectively. }
\label{fig:f4}
\end{figure}

\begin{figure}
\epsfig{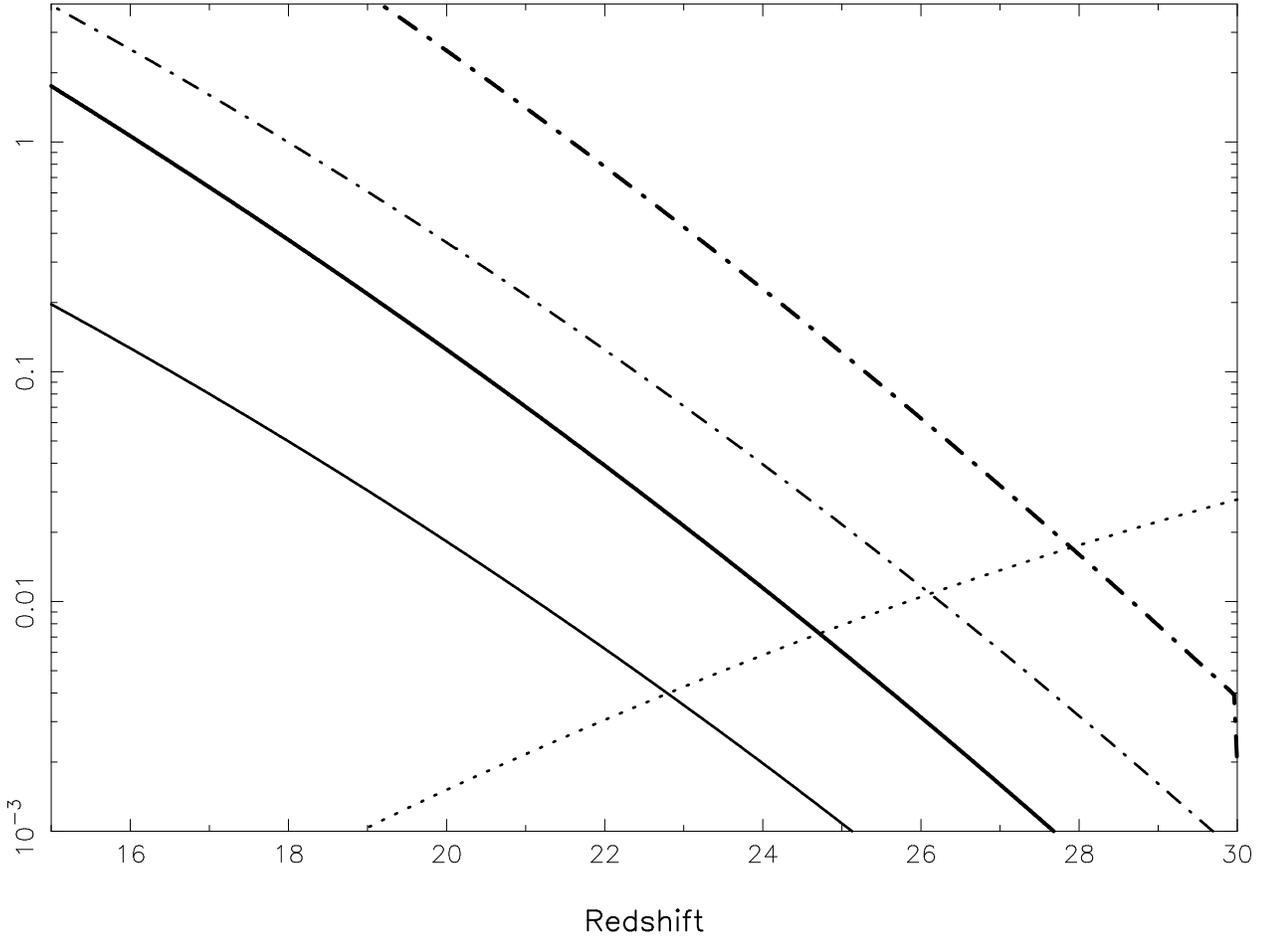}
\caption{The figure shows the influence of Lyman-$\alpha$ photons as 
compared to particle collision and CMBR on the spin temperature (Eq.~(\ref{eqts})) in regions that are not heated by x-ray radiation. Thin
and thick solid lines correspond to $y_\alpha T_K/T_{\rm CMBR}$ and $y_\alpha$
for $\dot N_\gamma(0) = 10^{50} \, \rm sec^{-1}$ and the ratio of Lyman-$\alpha$ to 
ionizing flux $A = 2$, respectively. The thin and thick dot-dashed lines 
correspond to the same quantities for $A = 40$. The dotted line 
represents $y_c$. } 
\label{fig:f5}
\end{figure}

\begin{figure}
\epsfig{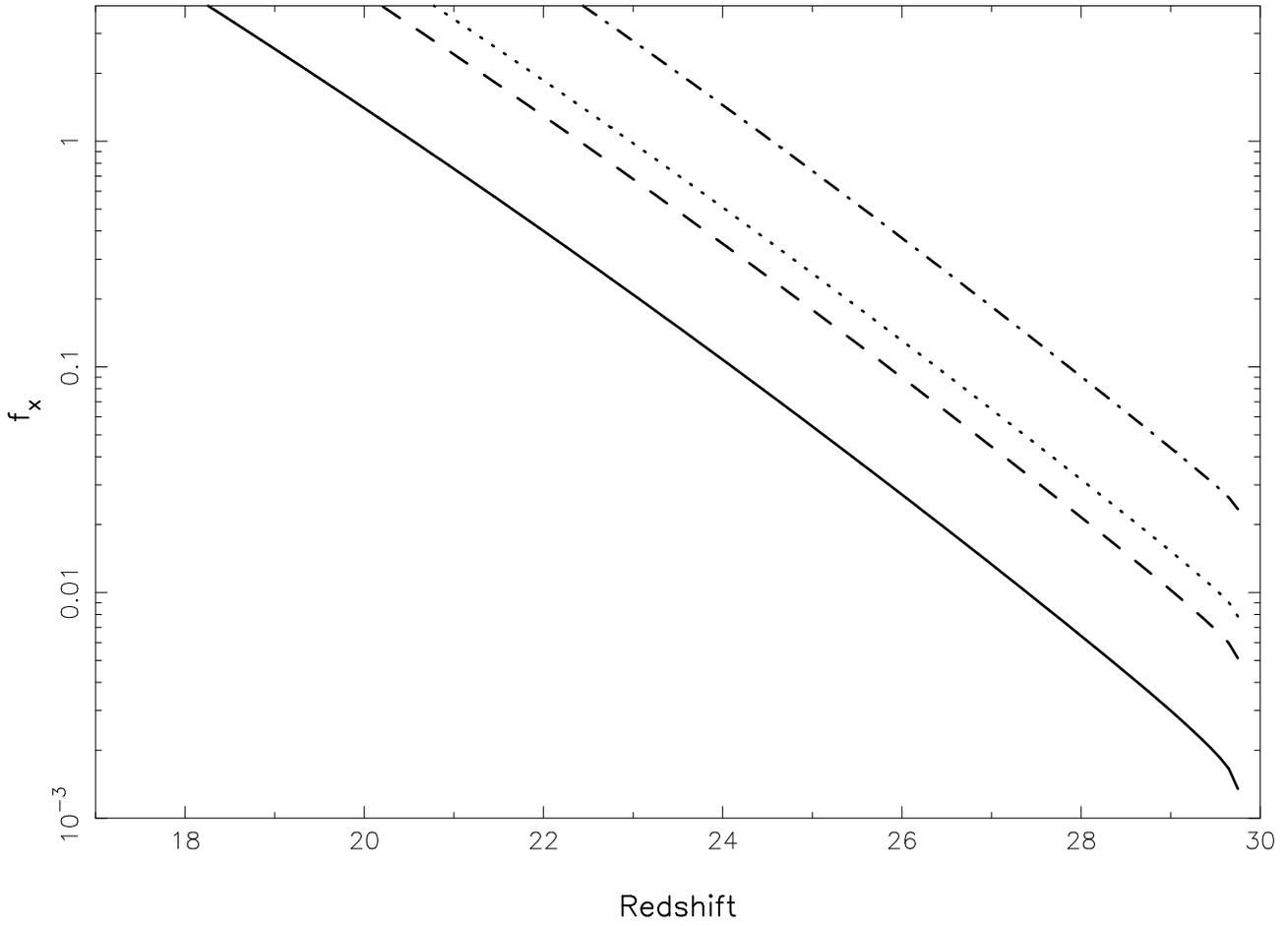}
\caption{The evolution of the fraction of the universe heated by
UV and soft xray, $f_{\rm x}$, is shown. All the models shown have
 $\dot N_\gamma(0) = 10^{50} \, \rm sec^{-1}$  and $C = 1$. Solid, dashed, and 
dot-dashed lines correspond to $q = 1$ and spectral indices $\alpha = \{1, 1.5, 2\}$, respectively. The dotted line corresponds to $q = 3$ and $\alpha = 1$}
\label{fig:f6}
\end{figure}

\begin{figure}
\epsfig{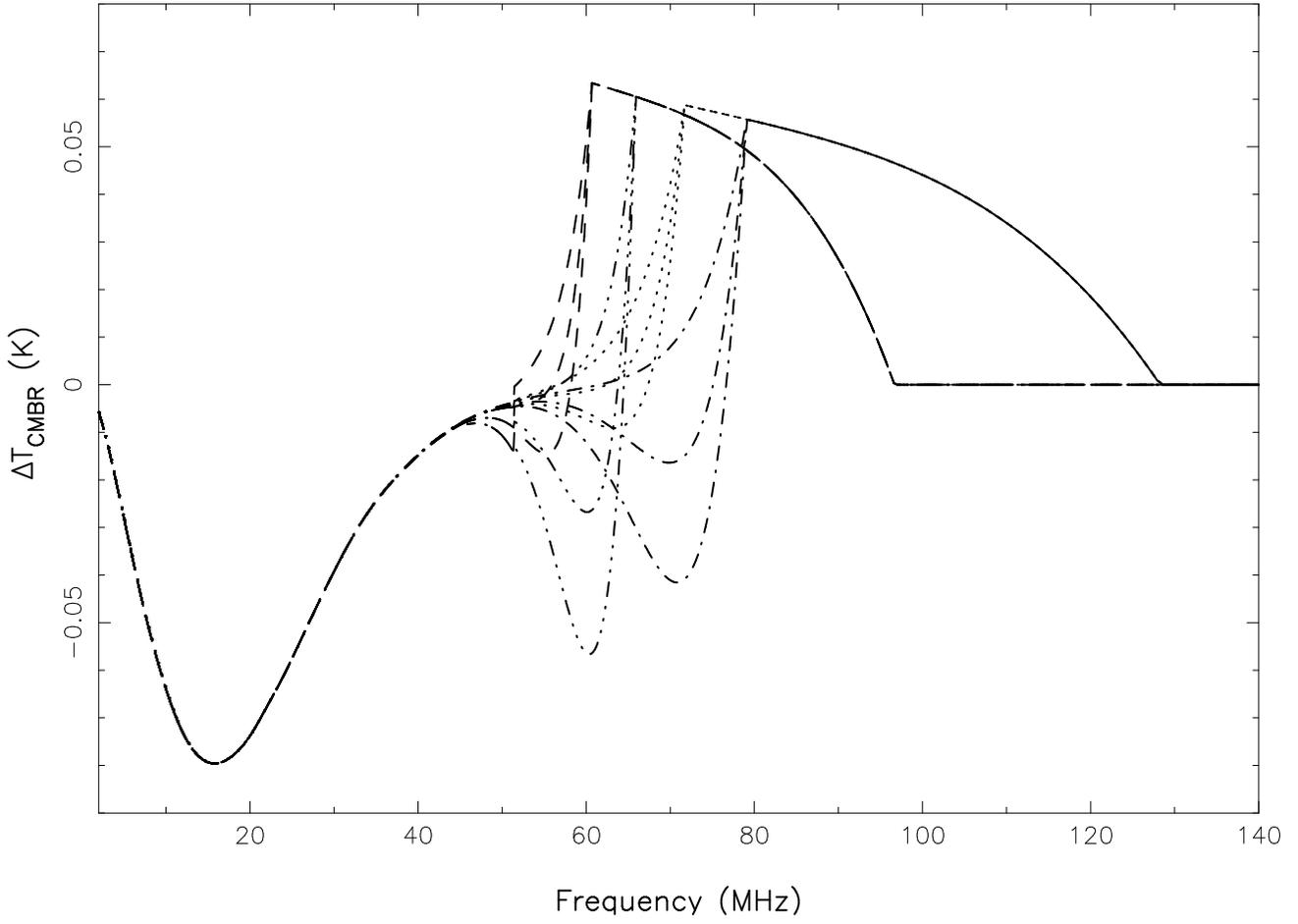}
\caption{The HI signal from pre-reionization and reionization epochs
is shown. The dashed and the dot-dot-dot-dashed  curves correspond
to an ionization history with $\dot N_\gamma(0) = 10^{50} \, \rm sec^{-1}$, $C = 1$. The dashed curves correspond to
soft xray spectral index $\alpha = 1.5$ and, from top to bottom, to ratio
of Lyman-$\alpha$ to ionizing flux ratios $\{2, 10, 20\}$. The dot-dot-dot-dashed curves correspond to $\alpha = 2$ with the same ratios of Lyman-$\alpha$
to ionizing flux. The dotted and dot-dashed curves correspond to an 
ionizing  history with $\dot N_\gamma(0) = 10^{49} \, \rm sec^{-1}$, $C = 1$; the values of $\alpha$ and ratio of 
Lyman-$\alpha$ and ionizing flux have the same values as the curves for larger
ionizing flux. }
\label{fig:f7}
\end{figure}

\begin{figure}
\epsfig{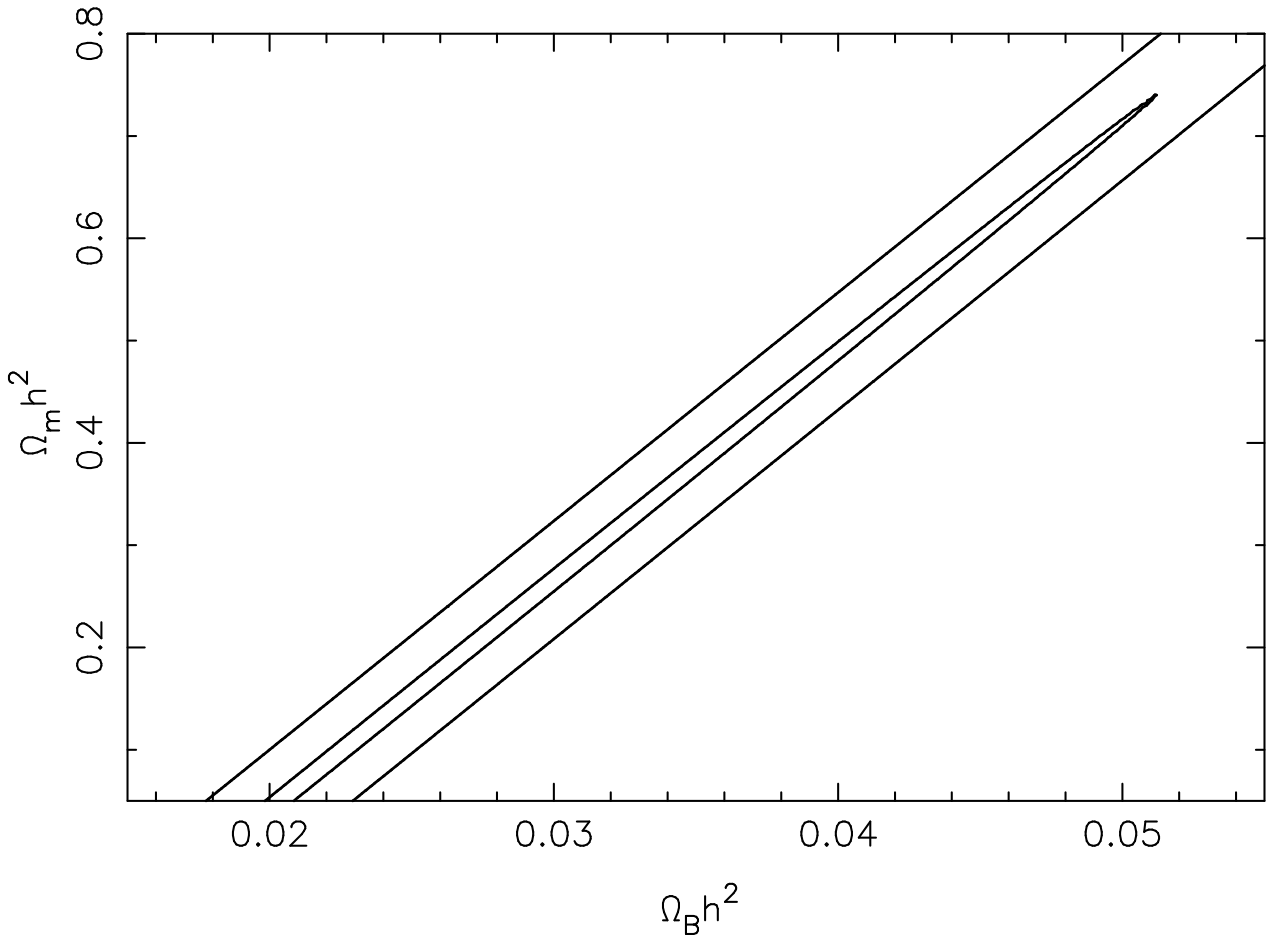}
\caption{For pre-reionization signal, 2-$\sigma$ contours in the plane of  
the cosmological parameters are shown. The inner and the outer
contour correspond to $S_{\rm rms} = \{1, 3\} \, \rm mK$, respectively}
\label{fig:f8}
\end{figure}

\begin{figure}
\epsfig{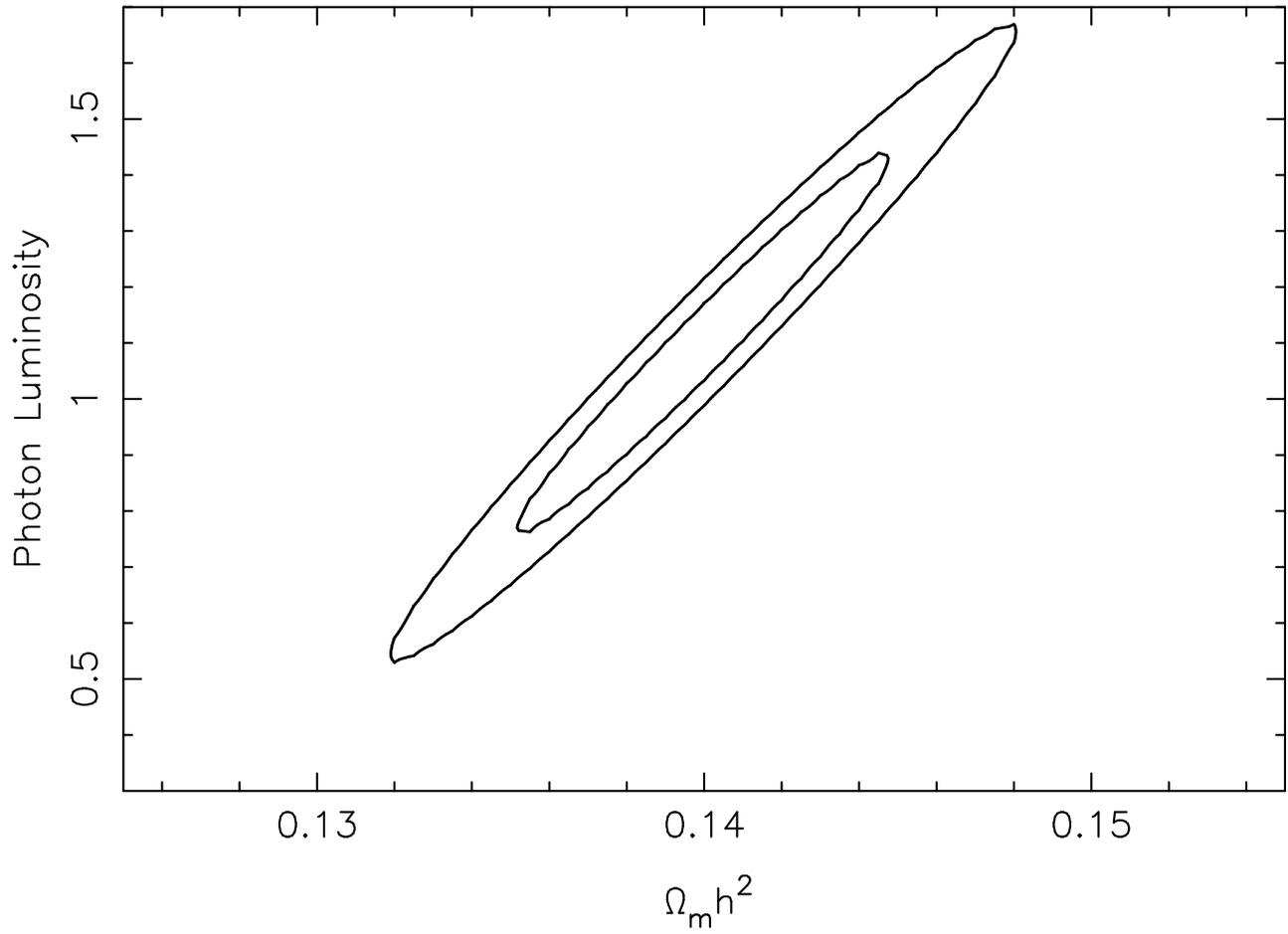}
\caption{For  the HI signal observed in emission, 
1~and 2-$\sigma$ contours in the plane of photon luminosity 
 $\dot N_\gamma(0)/10^{49}$ and $\Omega_mh^2$ are shown (for details
of the underlying model see text);  the RMS noise is 
 $S_{\rm rms} =  1 \, \rm mK$. }
\label{fig:f9}
\end{figure}

\begin{figure}
\epsfig{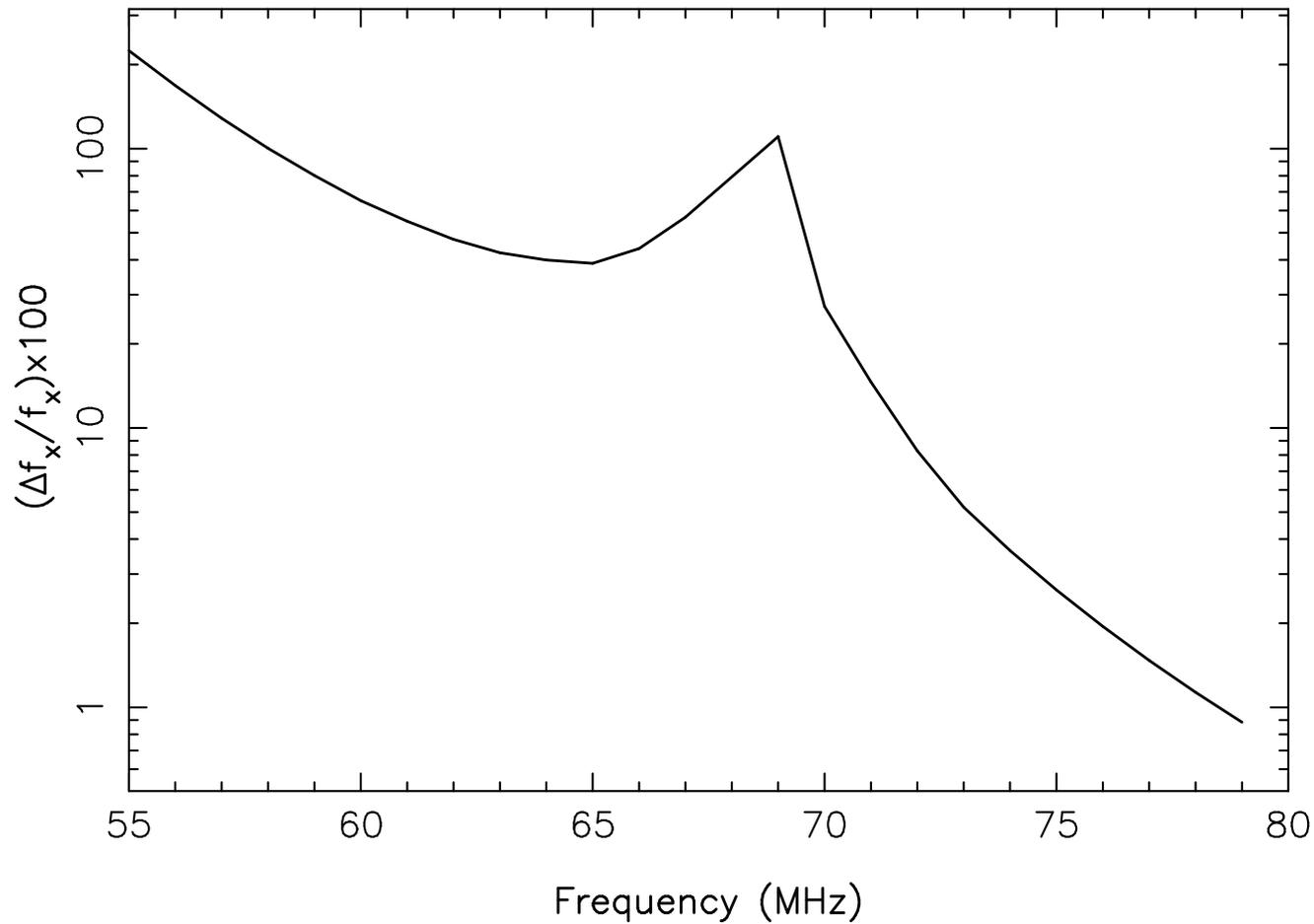}
\caption{The percentage error on the fraction of universe heated by 
soft x-ray above CMBR temperature 
 is plotted against the observing frequency. For details of the underlying 
model see text.}
\label{fig:f10}
\end{figure}

\end{document}